\documentclass[twocolumn]{aastex631}
\usepackage{CJK}
\usepackage[utf8]{inputenc}

\usepackage{upquote}

\begin{document}

\title{Spectropolarimetry of SN 2023ixf reveals both circumstellar material and an aspherical helium core}

\correspondingauthor{M. Shrestha}
\email{mshrestha1@arizona.edu}

\newcommand{\LCO}{\affiliation{Las Cumbres Observatory, 6740 Cortona Drive, Suite 102, Goleta, CA 93117-5575, USA}}
\newcommand{\UCSB}{\affiliation{Department of Physics, University of California, Santa Barbara, CA 93106-9530, USA}}
\newcommand{\KITP}{\affiliation{Kavli Institute for Theoretical Physics, University of California, Santa Barbara, CA 93106-4030, USA}}
\newcommand{\UCD}{\affiliation{Department of Physics and Astronomy, University of California, Davis, 1 Shields Avenue, Davis, CA 95616-5270, USA}}
\newcommand{\WIS}{\affiliation{Department of Particle Physics and Astrophysics, Weizmann Institute of Science, 76100 Rehovot, Israel}}
\newcommand{\OKC}{\affiliation{Oskar Klein Centre, Department of Astronomy, Stockholm University, Albanova University Centre, SE-106 91 Stockholm, Sweden}}
\newcommand{\OAPD}{\affiliation{INAF-Osservatorio Astronomico di Padova, Vicolo dell'Osservatorio 5, I-35122 Padova, Italy}}
\newcommand{\Caltech}{\affiliation{Cahill Center for Astronomy and Astrophysics, California Institute of Technology, Mail Code 249-17, Pasadena, CA 91125, USA}}
\newcommand{\GSFC}{\affiliation{Astrophysics Science Division, NASA Goddard Space Flight Center, Mail Code 661, Greenbelt, MD 20771, USA}}
\newcommand{\UMD}{\affiliation{Joint Space-Science Institute, University of Maryland, College Park, MD 20742, USA}}
\newcommand{\UCB}{\affiliation{Department of Astronomy, University of California, Berkeley, CA 94720-3411, USA}}
\newcommand{\TTU}{\affiliation{Department of Physics, Texas Tech University, Box 41051, Lubbock, TX 79409-1051, USA}}
\newcommand{\STScI}{\affiliation{Space Telescope Science Institute, 3700 San Martin Drive, Baltimore, MD 21218-2410, USA}}
\newcommand{\UT}{\affiliation{University of Texas at Austin, 1 University Station C1400, Austin, TX 78712-0259, USA}}
\newcommand{\IoA}{\affiliation{Institute of Astronomy, University of Cambridge, Madingley Road, Cambridge CB3 0HA, UK}}
\newcommand{\QUB}{\affiliation{Astrophysics Research Centre, School of Mathematics and Physics, Queen's University Belfast, Belfast BT7 1NN, UK}}
\newcommand{\IPAC}{\affiliation{Spitzer Science Center, California Institute of Technology, Pasadena, CA 91125, USA}}
\newcommand{\IPACMR}{\affiliation{IPAC, Mail Code 100-22, Caltech, 1200 E.\ California Blvd., Pasadena, CA 91125}}
\newcommand{\JPL}{\affiliation{Jet Propulsion Laboratory, California Institute of Technology, 4800 Oak Grove Dr, Pasadena, CA 91109, USA}}
\newcommand{\Southampton}{\affiliation{Department of Physics and Astronomy, University of Southampton, Southampton SO17 1BJ, UK}}
\newcommand{\LANL}{\affiliation{Space and Remote Sensing, MS B244, Los Alamos National Laboratory, Los Alamos, NM 87545, USA}}
\newcommand{\Tsinghua}{\affiliation{Physics Department and Tsinghua Center for Astrophysics, Tsinghua University, Beijing, 100084, People's Republic of China}}
\newcommand{\NAOC}{\affiliation{National Astronomical Observatory of China, Chinese Academy of Sciences, Beijing, 100012, People's Republic of China}}
\newcommand{\Itagaki}{\affiliation{Itagaki Astronomical Observatory, Yamagata 990-2492, Japan}}
\newcommand{\Einstein}{\altaffiliation{Einstein Fellow}}
\newcommand{\Hubble}{\altaffiliation{Hubble Fellow}}
\newcommand{\CfA}{\affiliation{Center for Astrophysics \textbar{} Harvard \& Smithsonian, 60 Garden Street, Cambridge, MA 02138-1516, USA}}
\newcommand{\UA}{\affiliation{Steward Observatory, University of Arizona, 933 North Cherry Avenue, Tucson, AZ 85721-0065, USA}}
\newcommand{\MPIA}{\affiliation{Max-Planck-Institut f\"ur Astrophysik, Karl-Schwarzschild-Stra\ss{}e 1, D-85748 Garching, Germany}}
\newcommand{\DSFP}{\altaffiliation{LSSTC Data Science Fellow}}
\newcommand{\HCO}{\affiliation{Harvard College Observatory, 60 Garden Street, Cambridge, MA 02138-1516, USA}}
\newcommand{\Carnegie}{\affiliation{Observatories of the Carnegie Institute for Science, 813 Santa Barbara Street, Pasadena, CA 91101-1232, USA}}
\newcommand{\TAU}{\affiliation{School of Physics and Astronomy, Tel Aviv University, Tel Aviv 69978, Israel}}
\newcommand{\Edinburgh}{\affiliation{Institute for Astronomy, University of Edinburgh, Royal Observatory, Blackford Hill EH9 3HJ, UK}}
\newcommand{\Birmingham}{\affiliation{Birmingham Institute for Gravitational Wave Astronomy and School of Physics and Astronomy, University of Birmingham, Birmingham B15 2TT, UK}}
\newcommand{\Bath}{\affiliation{Department of Physics, University of Bath, Claverton Down, Bath BA2 7AY, UK}}
\newcommand{\CTIO}{\affiliation{Cerro Tololo Inter-American Observatory, National Optical Astronomy Observatory, Casilla 603, La Serena, Chile}}
\newcommand{\Potsdam}{\affiliation{Institut f\"ur Physik und Astronomie, Universit\"at Potsdam, Haus 28, Karl-Liebknecht-Str. 24/25, D-14476 Potsdam-Golm, Germany}}
\newcommand{\INPE}{\affiliation{Instituto Nacional de Pesquisas Espaciais, Avenida dos Astronautas 1758, 12227-010, S\~ao Jos\'e dos Campos -- SP, Brazil}}
\newcommand{\UNC}{\affiliation{Department of Physics and Astronomy, University of North Carolina, 120 East Cameron Avenue, Chapel Hill, NC 27599, USA}}
\newcommand{\Ohio}{\affiliation{Astrophysical Institute, Department of Physics and Astronomy, 251B Clippinger Lab, Ohio University, Athens, OH 45701-2942, USA}}
\newcommand{\AAS}{\affiliation{American Astronomical Society, 1667 K~Street NW, Suite 800, Washington, DC 20006-1681, USA}}
\newcommand{\MMT}{\affiliation{MMT Observatory, PO Box 210065, University of Arizona, Tucson, AZ 85721-0065}}
\newcommand{\Geneva}{\affiliation{ISDC, Department of Astronomy, University of Geneva, Chemin d'\'Ecogia, 16 CH-1290 Versoix, Switzerland}}
\newcommand{\IUCAA}{\affiliation{Inter-University Center for Astronomy and Astrophysics, Post Bag 4, Ganeshkhind, Pune, Maharashtra 411007, India}}
\newcommand{\CMU}{\affiliation{Department of Physics, Carnegie Mellon University, 5000 Forbes Avenue, Pittsburgh, PA 15213-3815, USA}}
\newcommand{\NAOJ}{\affiliation{Division of Science, National Astronomical Observatory of Japan, 2-21-1 Osawa, Mitaka, Tokyo 181-8588, Japan}}
\newcommand{\IfA}{\affiliation{Institute for Astronomy, University of Hawai`i, 2680 Woodlawn Drive, Honolulu, HI 96822-1839, USA}}
\newcommand{\UCSC}{\affiliation{Department of Astronomy and Astrophysics, University of California, Santa Cruz, CA 95064-1077, USA}}
\newcommand{\Purdue}{\affiliation{Department of Physics and Astronomy, Purdue University, 525 Northwestern Avenue, West Lafayette, IN 47907-2036, USA}}
\newcommand{\Princeton}{\affiliation{Department of Astrophysical Sciences, Princeton University, 4 Ivy Lane, Princeton, NJ 08540-7219, USA}}
\newcommand{\Moore}{\affiliation{Gordon and Betty Moore Foundation, 1661 Page Mill Road, Palo Alto, CA 94304-1209, USA}}
\newcommand{\Durham}{\affiliation{Department of Physics, Durham University, South Road, Durham, DH1 3LE, UK}}
\newcommand{\JHU}{\affiliation{Department of Physics and Astronomy, The Johns Hopkins University, 3400 North Charles Street, Baltimore, MD 21218, USA}}
\newcommand{\Toronto}{\affiliation{David A.\ Dunlap Department of Astronomy and Astrophysics, University of Toronto,\\ 50 St.\ George Street, Toronto, Ontario, M5S 3H4 Canada}}
\newcommand{\Duke}{\affiliation{Department of Physics, Duke University, Campus Box 90305, Durham, NC 27708, USA}}
\newcommand{\NCU}{\affiliation{Graduate Institute of Astronomy, National Central University, 300 Jhongda Road, 32001 Jhongli, Taiwan}}
\newcommand{\Columbia}{\affiliation{Department of Physics and Columbia Astrophysics Laboratory, Columbia University, Pupin Hall, New York, NY 10027, USA}}
\newcommand{\Flatiron}{\affiliation{Center for Computational Astrophysics, Flatiron Institute, 162 5th Avenue, New York, NY 10010-5902, USA}}
\newcommand{\CIERA}{\affiliation{Center for Interdisciplinary Exploration and Research in Astrophysics and Department of Physics and Astronomy, \\Northwestern University, 1800 Sherman Avenue, 8th Floor, Evanston, IL 60201, USA}}
\newcommand{\GeminiNorth}{\affiliation{Gemini Observatory, 670 North A`ohoku Place, Hilo, HI 96720-2700, USA}}
\newcommand{\Keck}{\affiliation{W.~M.~Keck Observatory, 65-1120 M\=amalahoa Highway, Kamuela, HI 96743-8431, USA}}
\newcommand{\UW}{\affiliation{Department of Astronomy, University of Washington, 3910 15th Avenue NE, Seattle, WA 98195-0002, USA}}
\newcommand{\catalyst}{\altaffiliation{LSST-DA Catalyst Fellow}}
\newcommand{\USask}{\affiliation{Department of Physics \& Engineering Physics, University of Saskatchewan, 116 Science Place, Saskatoon, SK S7N 5E2, Canada}}
\newcommand{\Thacher}{\affiliation{Thacher School, 5025 Thacher Road, Ojai, CA 93023-8304, USA}}
\newcommand{\Rutgers}{\affiliation{Department of Physics and Astronomy, Rutgers, the State University of New Jersey,\\136 Frelinghuysen Road, Piscataway, NJ 08854-8019, USA}}
\newcommand{\FSU}{\affiliation{Department of Physics, Florida State University, 77 Chieftan Way, Tallahassee, FL 32306-4350, USA}}
\newcommand{\Melbourne}{\affiliation{School of Physics, The University of Melbourne, Parkville, VIC 3010, Australia}}
\newcommand{\ASTROthreeD}{\affiliation{ARC Centre of Excellence for All Sky Astrophysics in 3 Dimensions (ASTRO 3D)}}
\newcommand{\Stromlo}{\affiliation{Mt.\ Stromlo Observatory, The Research School of Astronomy and Astrophysics, Australian National University, ACT 2601, Australia}}
\newcommand{\NCPAS}{\affiliation{National Centre for the Public Awareness of Science, Australian National University, Canberra, ACT 2611, Australia}}
\newcommand{\TAMU}{\affiliation{Department of Physics and Astronomy, Texas A\&M University, 4242 TAMU, College Station, TX 77843, USA}}
\newcommand{\Mitchell}{\affiliation{George P.\ and Cynthia Woods Mitchell Institute for Fundamental Physics \& Astronomy, College Station, TX 77843, USA}}
\newcommand{\ESO}{\affiliation{European Southern Observatory, Alonso de C\'ordova 3107, Casilla 19, Santiago, Chile}}
\newcommand{\ICE}{\affiliation{Institute of Space Sciences (ICE, CSIC), Campus UAB, Carrer
de Can Magrans, s/n, E-08193 Barcelona, Spain}}
\newcommand{\IEEC}{\affiliation{Institut d'Estudis Espacials de Catalunya, Gran Capit\`a, 2-4, Edifici Nexus, Desp.\ 201, E-08034 Barcelona, Spain}}
\newcommand{\Warwick}{\affiliation{Department of Physics, University of Warwick, Gibbet Hill Road, Coventry CV4 7AL, UK}}
\newcommand{\Macquarie}{\affiliation{School of Mathematical and Physical Sciences, Macquarie University, NSW 2109, Australia}}
\newcommand{\AAARC}{\affiliation{Astronomy, Astrophysics and Astrophotonics Research Centre, Macquarie University, Sydney, NSW 2109, Australia}}
\newcommand{\Capodimonte}{\affiliation{INAF - Capodimonte Astronomical Observatory, Salita Moiariello 16, I-80131 Napoli, Italy}}
\newcommand{\INFNNapoli}{\affiliation{INFN - Napoli, Strada Comunale Cinthia, I-80126 Napoli, Italy}}
\newcommand{\ICRANet}{\affiliation{ICRANet, Piazza della Repubblica 10, I-65122 Pescara, Italy}}
\newcommand{\MSU}{\affiliation{Center for Data Intensive and Time Domain Astronomy, Department of Physics and Astronomy,\\Michigan State University, East Lansing, MI 48824, USA}}
\newcommand{\SETI}{\affiliation{SETI Institute,
339 Bernardo Ave, Suite 200, Mountain View, CA 94043, USA}}
\newcommand{\IAIFI}{\affiliation{The NSF AI Institute for Artificial Intelligence and Fundamental Interactions}}
\newcommand{\ANUC}{\affiliation{Department of Astronomy, AlbaNova University Center, Stockholm University, SE-10691 Stockholm, Sweden}}

\newcommand{\Konkoly}{\affiliation{Konkoly Observatory,  CSFK, Konkoly-Thege M. \'ut 15-17, Budapest, 1121, Hungary}}
\newcommand{\ELTE}{\affiliation{ELTE E\"otv\"os Lor\'and University, Institute of Physics, P\'azm\'any P\'eter s\'et\'any 1/A, Budapest, 1117 Hungary}}
\newcommand{\SZTE}{\affiliation{Department of Experimental Physics, University of Szeged, D\'om t\'er 9, Szeged, 6720, Hungary}}
\newcommand{\IdAlta}{\affiliation{Instituto de Alta Investigaci\'on, Sede Esmeralda, Universidad de Tarapac\'a, Av. Luis Emilio Recabarren 2477, Iquique, Chile}}
\newcommand{\Kavli}{\affiliation{Kavli Institute for Cosmological Physics, University of Chicago, Chicago, IL 60637, USA}}
\newcommand{\UofChicago}{\affiliation{Department of Astronomy and Astrophysics, University of Chicago, Chicago, IL 60637, USA}}
\newcommand{\Fermi}{\affiliation{Fermi National Accelerator Laboratory, P.O.\ Box 500, Batavia, IL 60510, USA}}
\newcommand{\Dartmouth}{\affiliation{Department of Physics and Astronomy, Dartmouth College, Hanover, NH 03755, USA}}
\newcommand{\Surrey}{\affiliation{Department of Physics, University of Surrey, Guildford GU2 7XH, UK}}
\newcommand{\NU}{\affiliation{Center for Interdisciplinary Exploration and Research in Astrophysics (CIERA) and Department of Physics and Astronomy, Northwestern University, Evanston, IL 60208, USA}}

\newcommand{\itagaki}{\affiliation{Itagaki Astronomical Observatory, Yamagata 990-2492, Japan}}

\newcommand{\DU}{\affiliation{Department of Physics \& Astronomy, University of Denver, 2112 East Wesley Avenue, Denver, CO 80208, USA}}

\newcommand{\sdss}{\affiliation{Department of Astronomy, San Diego State University, San Diego, CA 92182-1221, USA}}

\newcommand{\SDS}{\affiliation{Department of Astronomy \& Astrophysics, University of California, San Diego, 9500 Gilman Drive, MC 0424, La Jolla, CA 92093-0424, USA}}
\newcommand{\ARI}{\affiliation{Astrophysics Research Institute, Liverpool John Moores University, 146 Brownlow Hill, Liverpool L3 5RF, UK}}
\newcommand{\UofSh}{\affiliation{Department of Physics and Astronomy, University of Sheffield, Hicks Building, Hounsfield Road, Sheffield S3 7RH, UK}}
\newcommand{\RHUL}{\affiliation{Department of Physics, Royal Holloway - University of London, Egham, TW20 0EX, UK}}
\newcommand{\UofH}{\affiliation{Centre for Astrophysics Research, University of Hertfordshire, Hatfield, AL10 9AB, UK}}
\newcommand{\UVA}{\affiliation{Department of Astronomy, University of Virginia, Charlottesville, VA 22904, USA}}

\author[0000-0002-4022-1874]{Manisha Shrestha}
\UA
\author[0000-0003-4829-6499]{Sabrina DeSoto}
\DU

\author[0000-0003-4102-380X]{David J. Sand}
\UA

\author[0000-0002-3452-0560]{G. Grant Williams}
\MMT
\UA

\author[0000-0003-1495-2275]{Jennifer L. Hoffman}
\DU

\author[0000-0002-5083-3663]{Paul S. Smith}
\UA

\author[0000-0002-3375-3397]{Callum McCall}
\ARI
\author[0000-0003-0733-7215]{Justyn R. Maund}
\RHUL
\author[0000-0001-8397-5759]{Iain A Steele}
\ARI

\author[0000-0002-9133-7957]{Klaas Wiersema}
\UofH

\author[0000-0003-0123-0062]{Jennifer E. Andrews}
\GeminiNorth

\author[0000-0001-5510-2424]{Nathan Smith}
\UA

\author[0000-0002-8826-3571]{Christopher Bilinski}
\UA

\author{Peter Milne}
\UA

\author[0000-0002-4989-6253]{Ramya M Anche}
\UA
\author[0000-0002-4924-444X]{K. Azalee Bostroem}
\catalyst\UA

\author[0000-0002-0832-2974]{Griffin Hosseinzadeh}
\SDS
\author[0000-0002-0744-0047]{Jeniveve Pearson}
\UA

\author[0000-0001-7839-1986]{Douglas C. Leonard}
\sdss

\author[0000-0002-9454-1742]{Brian~Hsu}
\UA

\author[0000-0002-7937-6371]{Yize Dong \begin{CJK*}{UTF8}{gbsn}(董一泽)\end{CJK*}}
\CfA

\author[0000-0003-2744-4755]{Emily Hoang}
\UCD

\author[0000-0003-0549-3281]{Daryl Janzen}
\USask
\author[0000-0001-5754-4007]{Jacob E. Jencson}
\IPACMR
\author[0000-0001-8738-6011]{Saurabh W.\ Jha}
\Rutgers

\author[0000-0001-9589-3793]{M.~J. Lundquist}

\Keck
\author{Darshana Mehta}
\UCD

\author[0000-0002-7015-3446]{Nicol\'as Meza Retamal}
\UCD

\author[0000-0001-8818-0795]{Stefano Valenti}
\UCD

\author[0000-0003-4914-5625]{Joseph Farah}
\LCO 
\UCSB

\author[0000-0003-4253-656X]{D.\ Andrew Howell}
\LCO\UCSB
\author[0000-0001-5807-7893]{Curtis McCully}
\LCO

\author[0000-0001-9570-0584]{Megan Newsome}
\LCO 
\UCSB

\author[0000-0003-0209-9246]{Estefania Padilla Gonzalez}
\LCO
\UCSB
\author[0000-0002-7472-1279]{Craig Pellegrino}
\UVA
\author[0000-0003-0794-5982]{Giacomo Terreran}
\LCO 
\UCSB






\begin{abstract}

We present multi-epoch optical spectropolarimetric and imaging polarimetric observations of the nearby Type II supernova (SN) 2023ixf discovered in M101 at a distance of 6.85 Mpc. The first imaging polarimetric observations were taken +2.33 days (60085.08 MJD) after the explosion, while the last imaging polarimetric data points (+73.19 and +76.19 days) were acquired after the fall from the light curve plateau. At +2.33 days there is strong evidence of circumstellar material (CSM) interaction in the spectra and the light curve. A significant level of intrinsic polarization $p_r = 1.02\pm 0.07 \% $ is seen during this phase which indicates that this CSM is aspherical. We find that the polarization evolves with time toward the interstellar polarization level during the photospheric phase, which suggests that the recombination photosphere is spherically symmetric. There is a jump in polarization ($p_r =0.45 \pm 0.08 \% $ and $p_r =0.62 \pm 0.08 \% $) at +73.19 and +76.19 days when the light curve falls from the plateau. This is a phase where polarimetric data is sensitive to non-spherical inner ejecta or a decrease in optical depth into the single scattering regime. We also present spectropolarimetric data that reveal line (de)polarization during most of the observed epochs. In addition, at +14.50 days we see an ``inverse P Cygni" profile in the H and He line polarization, which clearly indicates the presence of asymmetrically distributed material overlying the photosphere. The overall temporal evolution of polarization is typical for Type II SNe, but the high level of polarization during the rising phase has only been observed in SN 2023ixf. 


\end{abstract}



\section{Introduction} \label{sec:intro}
Stars with masses greater than 8 $M_\odot$ end their lives as explosive core-collapse supernovae (CCSNe). 
Hydrogen-rich SNe, also known as Type II\footnote{Here we refer only to Types IIP and IIL, not Types IIn, IIb, or other peculiar subtypes.}, are the most common type of CCSNe \citep{Li_2011,Smith_2011} and are thought to be explosions of red supergiant (RSG) stars. Direct imaging of SN sites has confirmed the progenitors of type II SNe to be RSGs \citep[e.g.][]{Smartt_2015, vandyk_2017}. These transients provide a unique opportunity to probe massive star evolution and its impact on galaxy environments via the formation and distribution of heavy elements. However, there are still critical gaps in our knowledge. Direct observations of mass loss during the final stages, the last months to years before the explosion, are complicated and poorly understood \citep{Smith_2014}. Increasing evidence from spectroscopy in the form of narrow spectral emission lines from high-ionization states (``flash spectroscopy"; \citealt{GalYam_2014, Yaron_2017, Bruch_2021, Bostroem_2023_23ixf, Shrestha_2024_24ggi}) and from early light curve numerical modeling   \citep{Morozova_2017,Morozova_2018} points to brief but extreme precursor mass-loss events producing significant circumstellar material (CSM) in otherwise normal SNe II.

A powerful way to probe both the properties of the CSM  and the explosion mechanism is via polarimetric observations during the different phases of the SN evolution. The free electrons in the CSM surrounding the SN scatter the photons from the photosphere which produces polarization. The polarization of the scattered photons carries the imprint of the scattering medium. In the case of spectropolarimetry, the large-scale asymmetries of the scattering medium create continuum polarization, while line polarization effects are caused by smaller-scale line-specific phenomena such as clumps. Hence, polarimetry is effective in deducing the geometric structure of the SN ejecta as well as its CSM. For a detailed review of SN polarimetry, refer to \cite{Leonard_2006, Wang_2008, Chornock_2010, Nagao_2019, Bilinski_2023, Nagao_2023}. Polarimetry of SNe thus provides knowledge complementary to standard photometry and spectroscopy.  Furthermore, it can provide information about the geometry of unresolved distant objects that is not obtainable with other astronomical techniques. 

Observationally, a few Type II SNe have been observed with significant polarization \citep{Leonard_2001,Leonard_2006, Nagao_2019, Nagao_2023, Vasylyev_2023}. There is diversity in the observed polarization behavior with some Type II SNe showing low-level polarization during the plateau, and during the end of the plateau the polarization tends to peak \citep[e.g.,][]{Leonard_2006, Nagao_2023}. However, in the case of CSM-interacting Type II SNe, early-time polarization observations can help determine the geometry of the CSM, given that significant polarization is expected if the CSM is aspherical as seen for many type IIn SNe \citep[e.g.][]{Leonard_2000,Hoffman_2008_97eg,Bilinski_2023}. For the first time, significant polarization during the rising phase has been observed for a nearby Type II supernova, SN~2023ixf in M101 \citep{Vasylyev_2023, Maund_2023,Singh_2024_23ixfPol}.

SN~2023ixf was discovered on 2023-05-19 at 17:27:15 UTC (MJD~60083.72) in M101 at a distance of 6.85 $\pm$ 0.13 Mpc \citep{Riess_2022} by \cite{Itagaki_2023_23ixf_discovery}. The J2000 coordinates of the SN are RA = 14:03:38.562 and Dec.\ = +54:18:41.94 \citep{Jones_2023} as presented in \autoref{tab:prop}. A spectrum taken the same day (MJD 60083.933) classified it as a young type II SN \footnote{Here we do not differentiate between Types IIP and IIL.} with flash ionization features \citep{Perlely_2023_23ixfClassification}. The early photometric and spectroscopic evolution of SN~2023ixf has been studied comprehensively \citep[e.g.][]{Bostroem_2023_23ixf,Hosseinzadeh_2023_23ixf,Jacobson-Galan_2023,Hiramatsu_2023,Smith_2023,Li_2024_23ixf,Zhang_2023_23ixf,Singh_2024_23ixfPol,Zimmerman_2024}. \citet{Hosseinzadeh_2023_23ixf} used early light curve data to infer an explosion epoch of MJD 60082.75, which we adopt for this paper\footnote{All the phases quoted in this paper are with respect to this explosion epoch unless otherwise noted}.   We present the general properties of SN~2023ixf in \autoref{tab:results}. 

In addition, early spectropolarimetric observations of SN~2023ixf showed significant polarization ($P_r =1.02 \pm 0.08 \%$) at the earliest times (+1.4d, +2.5d), which began to decline at +3.5d \citep{Vasylyev_2023}.  The flash ionization features along with the behavior of the multiwavelength early light curves and polarization point to CSM interaction. Here we present optical imaging polarimetry and spectropolarimetry spanning from +2.33 days to +76.19 days after explosion. Note that a subset of the data presented herein was presented by \citet{Singh_2024_23ixfPol}. 
In this paper, we build on this work by presenting $R$-band imaging polarization values from +2.33 to +76.19 days and spectropolarimetric data with observation dates that span the gap between the results presented in \citet{Vasylyev_2023} and \citet{Singh_2024_23ixfPol}. These new data provide us with a complete picture of SN~2023ixf's geometrical evolution.

In this paper, we present linear spectropolarimetric and imaging polarimetry\footnote{We only consider linear polarization throughout this work.} data for SN~2023ixf, ranging from the brightening phase to the fall from the plateau. The paper is organized as follows: first, we describe the observations and data reduction in \autoref{sec:obs}. This is followed by the calculation of interstellar polarization (ISP) and results from imaging polarimetric and spectropolarimetric observations in \autoref{sec:ip}. 
We compare the results from SN~2023ixf with other SNe in \autoref{sec:comp}. Finally, we discuss and conclude in \autoref{sec:conclusions}.

\section{Observations and Data reduction} \label{sec:obs}
In this section, we present the data reduction techniques we followed for the MOPTOP imaging polarimetry and SPOL spectropolarimetry data. 
\begin{table}
 \caption{Properties of SN~2023ixf} \label{tab:results}
 \begin{tabular}{ll}
    \hline
    Parameter & Value \\
    \hline
    R.A. (J2000) & 14:03:38.562 \\
    Dec. (J2000) & $+$54:18:41.94 \\
    Explosion Epoch (MJD)\tablenotemark{a} & 60082.75 \\
    Distance modulus ($\mu$)\tablenotemark{b} & 29.178 $\pm$ 0.041 mag\\
    Peak Magnitude ($r_{\mathrm{max}}$) & $-18.07 \pm 0.04$ mag\\
    Time of $r_{\mathrm{max}}$ (MJD) & 60094.31 \\
    $T_{pt}$\tablenotemark{c} & 81.50$\pm$ 0.11  days \\
    
    \hline
 \end{tabular}
 \label{tab:prop}
 \tablenotetext{a}{value from \citet{Hosseinzadeh_2023_23ixf}}
 \tablenotetext{b}{value from \citet{Riess_2022}}
 \tablenotetext{c}{Estimated value of the time of drop off of the optical light-curve plateau, from \citet{hsu2024yearsn2023ixfbreaking}}

\end{table}

\subsection{MOPTOP data reduction}\label{subsec:obs_img}
We followed up SN~2023ixf using the imaging polarimeter MOPTOP\footnote{\url{https://telescope.livjm.ac.uk/TelInst/Inst/MOPTOP/}} \citep{Jermak_2016, Jermak_2018, Shrestha_2020} mounted on the 2-m Liverpool Telescope (LT) \citep{Steele_2004}. MOPTOP is a dual beam, dual camera imaging polarimeter with 7 arcmin $\times$ 7 arcmin field of view designed for maximum sensitivity for rapidly fading sources such as gamma-ray bursts and SNe. The instrument is equipped with $B$, $V$, $R$, and $I$ band filters. Imaging polarimetric observations of SN~2023ixf in $B$, $V$, $R$, and $I$ filters began on 60085.08 MJD (+2.33 days) and ended on 60158.95 MJD (+76.19 days). However, we only present $R$-band data in this paper because we have a complete temporal sequence for this filter. 

MOPTOP produces 16 different images from each camera, where each frame corresponds to 22.5$\degr$ rotation of the wave plate. The data reduction pipeline running at the telescope performs dark subtraction and flat-fielding. We manually perform aperture photometry to extract background-subtracted counts of the source. These counts are then converted to fractional Stokes parameter $q_{obs}$, $u_{obs}$ and their error via the `two-camera' technique as described in \cite{Shrestha_2020} (Eq. 14, 17 in the paper).  We use instrumental Stokes $q_{inst}$ and $u_{inst}$ values of MOPTOP to correct for instrumental effects and obtain the intrinsic values, given by $q_c = q_{obs} - q_{inst}$ and  $u_c = u_{obs} - u_{inst}$. These values are used to calculate polarization $p = \sqrt{q_c^2+u_c^2}$ and position angle $PA =\frac{1}{2} \arctan\left ( \frac{u_c}{q_c} \right)$. There is a polarization bias due to noise in $q$ and $u$ as $p$ is always a positive quantity. We correct for this using the technique presented in
\citet{Plaszczynski_2014}. Finally, we subtract Stokes $q$ and $u$ introduced by ISP. The calculation of ISP contribution is described in detail in \autoref{subsec:isp}.

\subsection{SPOL data reduction} \label{subsec:obs_spol}

We obtained optical (4500--7500 \AA) spectropolarimetric data for SN~2023ixf using the CCD Imaging/Spectropolarimeter (SPOL; \citealt{Schmidt_1992b}) mounted on  Steward Observatory's 2.3-m Bok telescope (Kitt Peak, AZ) between MJD 60095.15 (+12.40 days) and MJD 60114.32 (+31.57 days; \autoref{tab:pol}). To constrain the instrumental polarization we observed multiple polarized standards (HD155528, Hiltner 960, and VI Cyg 12) and unpolarized standards (BD+28 4211 and Wolf 1346). We find a low level of instrumental polarization (less than 0.1\%) from the observed unpolarized standard stars. Polarized standard star observations were used to calculate the offset in position angle. The data obtained by SPOL were reduced using custom IRAF routines; further details can be found in \citealt{Milne_2017} and \citealt{Bilinski_2023}. In short, the routine first flat-fields and bias-subtracts each image, which is wavelength calibrated based on He, Ne, and Ar lamp observations taken on the same night. From this, Stokes $q = Q/I$ and $u = U/I$ are extracted. We bin the Stokes parameters to 20 \AA~(similar to the native resolution of the SPOL setup) to improve the signal-to-noise ratio. These Stokes parameters are used to calculate the total polarization as $p = \sqrt{q^2+u^2}$. 

As a proxy for $R$ band imaging data from SPOL observations, we perform an error-weighted average of $q$ and $u$ for all the SPOL data. For consistency with the MOPTOP bias correction method, we correct for polarization bias introduced by noise in Stokes $q$ and $u$ via the method given by  \citet{Plaszczynski_2014} for the $R$ band calculations. Observations of a large data set of polarized and unpolarized standard stars, taken for over a decade with SPOL, show a standard deviation of 0.05$\%$. Thus, with these statistics, we adopt the systematic uncertainty of $q = 0.05\%$ and $u = 0.05\%$, as the minimum uncertainty for all the polarization measurements. We subtract ISP contributions as calculated in \autoref{subsec:isp} from these data. We also calculate position angle ($PA$) following the equation $PA =\frac{1}{2} \arctan\left ( \frac{u}{q} \right)$ and calculate error in $PA$ by propagating the error. 

We note that the SPOL data from 60109.16 to 60114.32 MJD were previously presented by \citet{Singh_2024_23ixfPol}. We include them in our analysis, as they have been uniformly reduced with the rest of the SPOL dataset presented herein.  Due to the relatively low S/N ratio, we only analyze the continuum polarization from these spectra, which we calculate via the error-weighted mean of $q$ and $u$ in the 6000--7000 \AA\ range as a proxy for $R$ band data. We use these integrated SPOL data to augment the imaging polarimetry sequence described in \autoref{subsec:obs_img}. For the spectropolarimetric data presentation, we refer the reader to \citet{Singh_2024_23ixfPol}.

\begin{table*}
 \centering
 \caption{Log of Polarimetric Observations of SN 2023ixf}
 \begin{tabular}{c c c c   c c c c c}
    \hline
     Date (UT) & MJD &  Phase (d)\tablenotemark{a} & Telescope & Instrument & Exposure time (s) & Airmass & $p\%$ ($R$ band) & $PA\degr$ ($R$ band) \\
    \hline
    
    2023-05-21 01:55:12 & 60085.08 & 2.33 &   LT & MOPTOP & 3 $\times$ 16 $\times$ 0.4  & 1.25 & 1.02 $\pm$ 0.07 &152 $\pm$ 4 \\
    2023-05-31 03:36:00 & 60095.15 &12.40&Bok& SPOL& 11 $\times$ 960    & 1.11 &0.26 $\pm$ 0.07& 59 $\pm$16\\
    2023-06-01 03:36:00 &60096.15 &13.40&Bok& SPOL&9 $\times$ 720     &  1.1& 0.23 $\pm$ 0.07& 61 $\pm$ 19 \\
    2023-06-02 06:00:00 & 60097.25 &14.50&Bok& SPOL& 12 $\times$ 720    & 1.11 & 0.21 $\pm$ 0.07& 65 $\pm$ 20\\
    2023-06-03 03:36:00 & 60098.15 &15.40&Bok& SPOL&10 $\times$ 720   & 1.11 & 0.21 $\pm$ 0.07& 67 $\pm$ 21\\
    2023-06-04 03:36:00 &60099.15 &16.40&Bok& SPOL& 10 $\times$ 720    &1.1  &0.19 $\pm$ 0.07& 70 $\pm$ 22\\
    2023-06-05 03:50:24 & 60100.16 &17.41&Bok& SPOL& 11 $\times$ 720    & 1.09 & 0.16 $\pm$ 0.07 & 73 $\pm$ 28 \\
    2023-06-14\tablenotemark{b} 03:50:24 & 60109.16 &26.41&Bok& SPOL&6 $\times$ 720   & 1.08 & 0.08 $\pm$ 0.07 &146 $\pm$ 68\\
    2023-06-15\tablenotemark{b} 03:36:00 & 60110.15 &27.40&Bok& SPOL& 3 $\times$ 720   & 1.08 & 0.08 $\pm$ 0.07 &150 $\pm$ 63\\
    2023-06-16\tablenotemark{b} 03:36:00 & 60111.15 &28.40&Bok& SPOL&5 $\times$ 720   &1.08  & 0.12 $\pm$ 0.07 &155 $\pm$ 68\\
    2023-06-18\tablenotemark{b} 03:36:00 & 60113.15 &30.40&Bok& SPOL& 4 $\times$ 720  &1.08  & 0.17 $\pm$ 0.07 &162 $\pm$ 38\\
    2023-06-19\tablenotemark{b} 07:40:48 &60114.32 &31.57&Bok& SPOL&5 $\times$ 720  & 1.48 & 0.19 $\pm$ 0.07 &166 $\pm$ 25\\
    2023-07-30 22:33:36 &60155.94 &73.19& LT & MOPTOP&5 $\times$ 16 $\times$ 0.4  & 1.5 & 0.45 $\pm$ 0.08& 22 $\pm$ 10\\
    2023-08-02 22:48:00 &60158.95 &76.19& LT & MOPTOP& 5 $\times$ 16 $\times$ 0.4 & 1.7& 0.62 $\pm$ 0.08& 178 $\pm$ 7\\
    
    \hline
 \end{tabular}
 \tablenotetext{a}{We list phase in days since the explosion epoch of MJD 60082.75.}
 \tablenotetext{b}{This set of data has been previously published in \citet{Singh_2024_23ixfPol}.}
 
 \label{tab:pol}
\end{table*}


\section{Polarimetry} \label{sec:ip}
In this section, we first calculate the polarization contribution from the interstellar medium using two different techniques described in detail in \autoref{subsec:isp}. Then we present our results from spectropolarimetric observations and discuss the behavior of the line polarization. Finally, we present the results from imaging polarimetry from MOPTOP and $R$-band continuum polarization estimates from SPOL data.  

\begin{figure}
    \centering
    \includegraphics[width=\columnwidth]{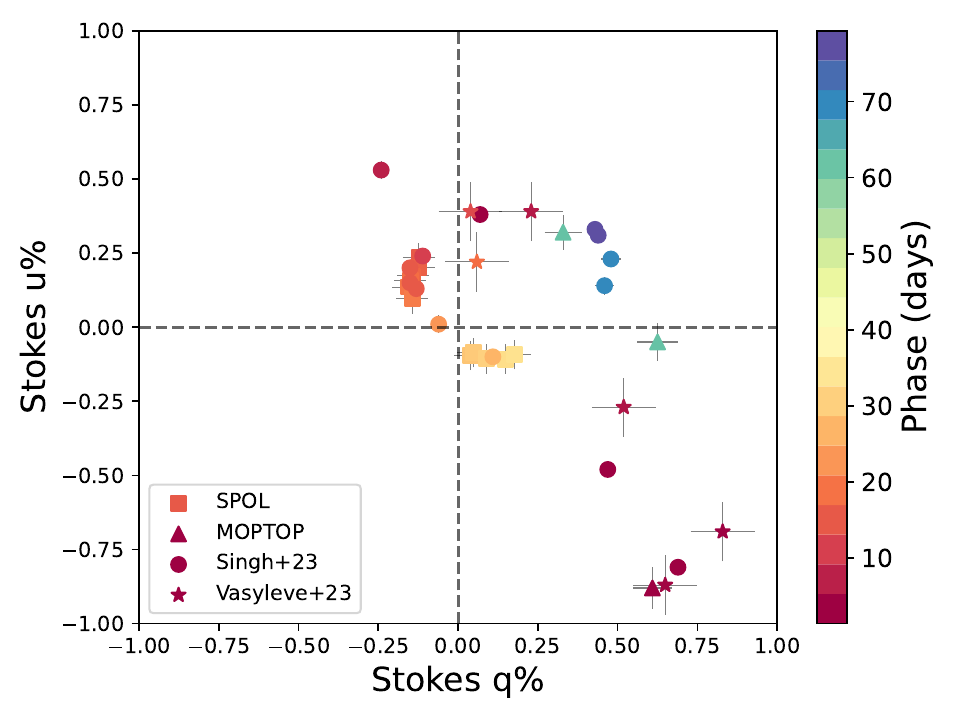}
    \caption{Evolution of $R$-band Stokes $q$ and $u$ with phase for SN 2023ixf after ISP correction (\S~\ref{subsec:isp})}. The points include data from MOPTOP, SPOL,and data from \citet{Vasylyev_2024} and \citet{Singh_2024_23ixfPol}. For SPOL data we calculated the continuum polarization in the 6000--7000 \AA~range (\autoref{subsec:obs_spol}). 
    \label{fig:qu_continuum}
\end{figure}

\subsection{Interstellar polarization} \label{subsec:isp}
Polarization measurements of a SN are contaminated by ISP due to the intervening dust in the Milky Way and the host galaxy. To extract the intrinsic polarization of SN~2023ixf, we employ two different techniques.

First, we use the dust extinction for the Milky Way and the host galaxy to estimate an upper limit in ISP. For SN~2023ixf, the dust extinction from the Milky Way is $E(B-V)_{MW}$ =  0.0074 mag \citep{Schlafly_2011}. To calculate the host galaxy extinction, \cite{Smith_2023} used high-resolution spectra to measure the equivalent width of the Na I D absorption lines at the redshift of SN~2023ixf and found $E(B-V)_{Host}$ = 0.031 mag\footnote{There is significant uncertainty in the relation from \citet{Poznanski_2012}}. Thus, the total extinction in the direction of SN~2023ixf is $E(B-V)$ =  0.0384 mag. Since the extinction value is greater than 0.01 mag as noted in \citet{Skalidis_2018}, the upper limit in polarization in the $V$ band induced by ISP (corresponding to the case in which the MW and host ISP align in angle) can be calculated using $p^{V}(ISP) < 9 E(B-V) \%$ \citep{Serkowski_1975}, assuming the dust in the host galaxy behaves similarly to the Milky Way dust.  We caution that for low extinction environments, there are sightlines that show much higher levels of linear polarization than expected from this scaling relation \citep[see][]{Skalidis_2018}. With $E(B-V)=$ 0.0384 mag, we calculate the upper limit of the $V$-band interstellar polarization as $p^{V}(ISP) < 0.35\%$. 
This is consistent with the values of 0.35$\%$ and 0.37$\%$ calculated by \citet{Vasylyev_2023} and \citet{Singh_2024_23ixfPol}, respectively. 

To calculate the ISP behavior as a function of wavelength, we utilize $p/p_{max} = \exp[-K\ln^2(\lambda_{max}/\lambda)]$, where $p_{max}$ is equivalent to  $p^{V}(ISP)$, the maximum polarization produced at the wavelength $\lambda_{max} = 5300$ \AA~(an approximation for the $V$ band). The value of $\lambda_{max} = 5300$ \AA~ in \citet{Serkowski_1975} is valid for dust grains similar to Milky Way dust, and we assume an $R_V = 3.1$ extinction law \citep{Cardelli_1989}. In addition, 
$K$ was originally a constant equal to 1.15 \citep{Serkowski_1975}, but \cite{Wilking_1982} later modified it to be $K = -0.10 + 1.86\lambda_{max}$, which we adopt in this paper. Here $\lambda_{max}$ is in units of microns, i.e., 0.53 $\mu$m. 

For the second method, we utilize the polarization associated with the strongest emission line (H$\alpha$), as the greater line flux is assumed to dilute and depolarize the continuum \citep[e.g.][]{Leonard_2000_halpha_isp,Leonard_2001,Nagao_2019,Leonard_2021,Singh_2024_23ixfPol}. The spectropolarimetry data from SPOL between 2023-06-14 and 2023-06-19 show a prominent H$\alpha$ feature. We calculated error-weighted averages of the Stokes $q$ and $u$ values from five different bins ranging from 100 to 200 $\AA$ in wavelength centered on the H$\alpha$ peak. We found similar values (within standard deviations of 0.007 and 0.02 for Stokes $q$ and $u$, respectively) as we changed the binning range from 100 \AA~to 200 \AA. From this method we calculated $q_{ISP} = -0.08 \pm 0.05 \%$ and $u_{ISP} = 0.10 \pm 0.05 \%$ centered at H$\alpha$ peak (an approximation for the $R$ band). These values are consistent with the ISP estimate calculated by \citet{Singh_2024_23ixfPol}. Our resulting $p_{ISP} =  0.13 \%$ value from this method is consistent with $p^{V}(ISP) < 0.35\%$ calculated from the first method. We follow \citet{Serkowski_1975} prescription to calculate the ISP contribution at different wavelengths using $p/p_{max} = \exp[-K\ln^2(\lambda_{max}/\lambda)]$ with $\lambda_{max} = 5300 \AA$, $K = -0.10 + 1.86\lambda_{max}$, and $p_{max} = 0.14\%$.
We subtract the $q_{ISP}$ and $u_{ISP}$ calculated for each filter from our observed data. We quote and display the ISP-corrected values throughout this work unless otherwise stated.

In \autoref{fig:qu_continuum}, we present the ISP-corrected continuum polarization measurements we obtained from MOPTOP, SPOL, \citet{Singh_2024_23ixfPol}, and \citet{Vasylyev_2023} data. While most of these continuum data show intrinsic polarization values, the last few data points during the plateau phase (from +26.41 to 31.57 days) lie very close to the origin of the $q-u$ plane ($\sim 0.05 \%$ for both $q$ and $u$) as expected during the photospheric phase. 

\subsection{Continuum polarization \& imaging polarimetry}\label{subsec:img}

\begin{figure*}
    \centering
    \includegraphics[width=0.75\textwidth]{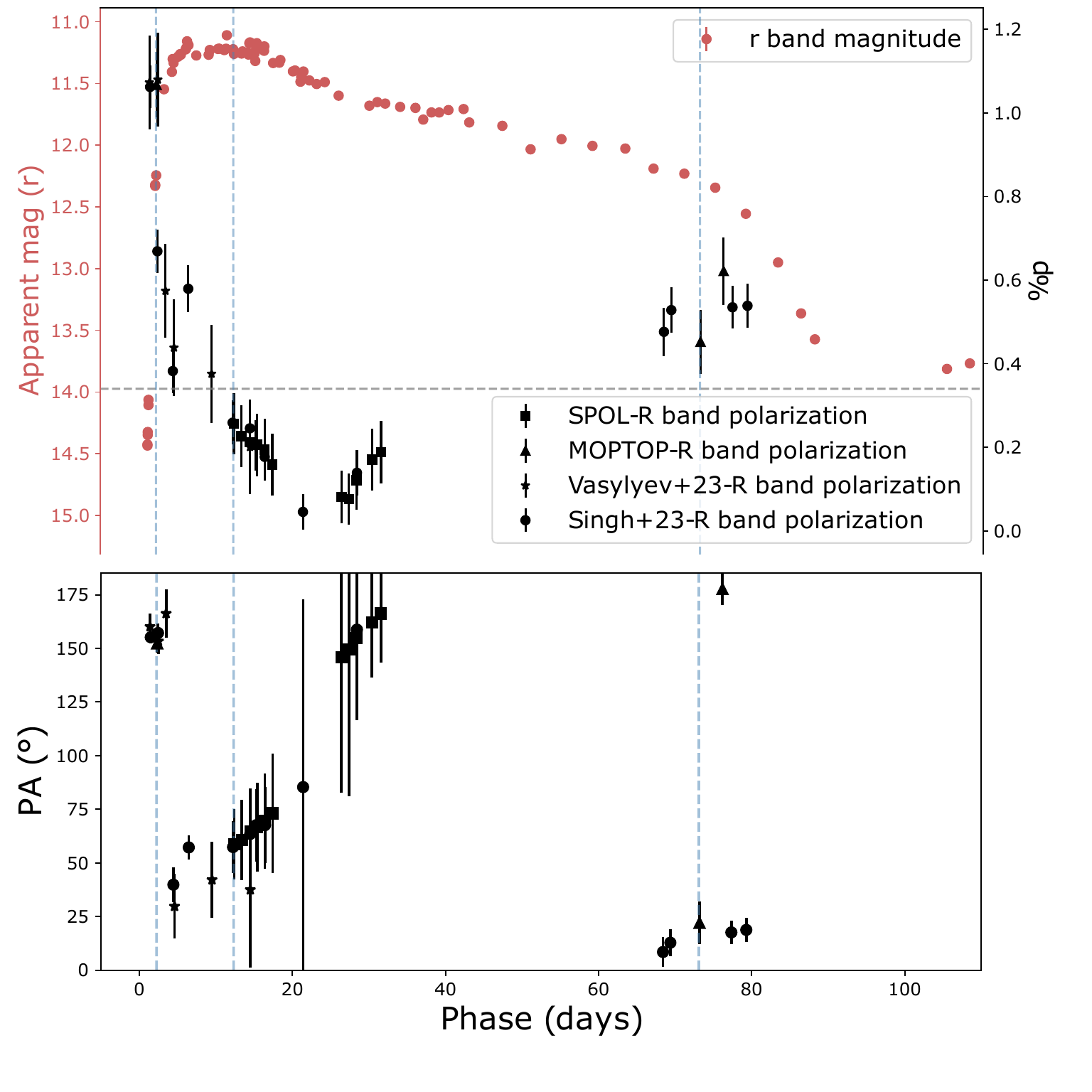}
    \caption{(top) $r$-band apparent magnitude (from Las Cumbres Observatory) and $R$-band polarization with respect to days since the explosion. The light dashed gray line is the maximum ISP in the $R$ band as discussed in \autoref{sec:ip}. The triangle data points are from MOPTOP observations and the square symbols are continuum polarization calculated from our SPOL data in the 6000--7000 \AA\ range. Circle and star points represent data from \citet{Singh_2024_23ixfPol} and \citet{Vasylyev_2023}, respectively.} Polarization detections during the initial rising phase and fall from the plateau suggest asphericity in the CSM and He core, respectively. (bottom) $PA$ with respect to phase for $R$-band data. In both panels, the dashed blue lines represent epochs that are presented in the \autoref{fig:sketch} sketch. 
    \label{fig:magpol}
\end{figure*}

The calculated continuum polarization and $PA$ from our observations are presented in \autoref{tab:pol}. To see how changes in the polarization correspond to changes in the light curve, we obtained an $r$-band light curve from the Sinistro cameras on Las Cumbres Observatory's robotic 1m telescopes \citep{Brown_2013} as part of the Global Supernova Project (GSP) collaboration \citep{Howell_2017}, which was published in \citet{hsu2024yearsn2023ixfbreaking}. The $r$-band magnitude peaks at $-18.07$ mag at 60094.31 MJD. 
In \autoref{fig:magpol}, we present the evolution of the $R$-band polarization with time, 
along with the $r$-band light curve (top panel). 
The first and the last two polarimetric data points at phases +2.33, +73.19, and +76.19 days are from MOPTOP on LT, while the data between these points are from SPOL.
The bottom panel of the figure shows the corresponding $PA$ values for the polarization measurements above. The dashed gray line represents the ISP upper limit calculated in \autoref{subsec:isp}.  

\begin{figure*}
    \centering
    \includegraphics[width=\textwidth]{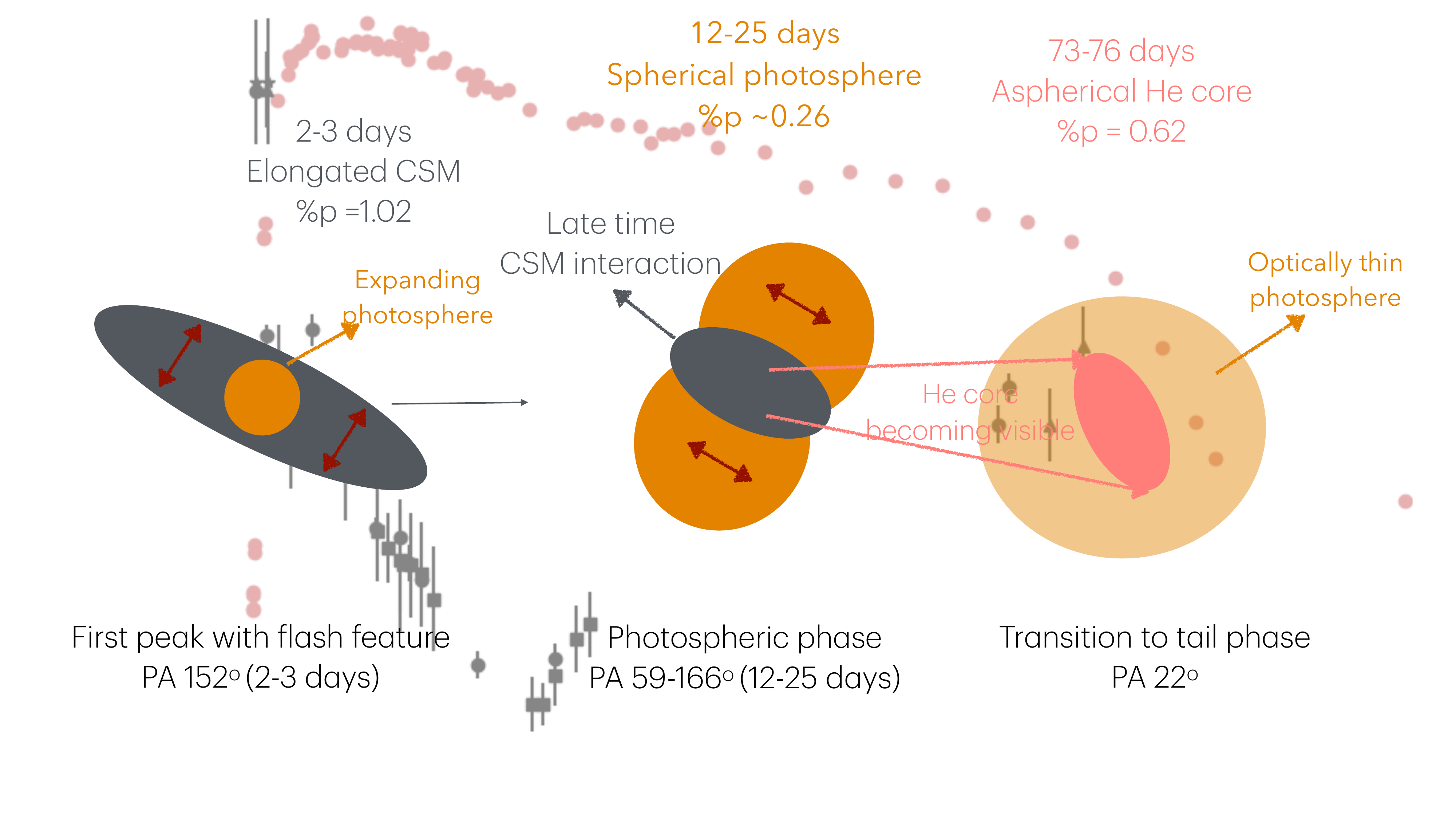}
    \caption{Simplified cross-section sketch with relative sizes not to scale of the evolution of CSM and ejecta geometry of SN~2023ixf based on our imaging and spectropolarimetric data. The background pink data points show the $r$ band apparent magnitude, and the black data points are imaging polarimetric data from \autoref{fig:magpol}. The shape of the scattering regions has been exaggerated for better visual representation. (Left) Elongated CSM (gray region) surrounding the photosphere (orange) with a scattered vector with (red) $PA$ of 153.9$\degr$ for the observation 2-3 days after the explosion. During this phase, there is strong evidence of CSM interaction from the spectra and the light curve. Hence, electron scattering from disk-shaped CSM during this phase produces polarization of 1.02$\%$ (\autoref{subsec:img}). (Middle) Our first spectropolarimetry data were taken +12.40 days after the explosion, featuring high continuum polarization and inverse P-Cygni line features. The continuum polarization is now due to the expanding photosphere (orange), which was constrained by the early CSM but gradually becomes more spherical during the plateau phase as it sweeps away the CSM. The inverse P-Cygni features are likely due to intervening material from the remnant CSM (gray region)  (\autoref{subsec:line}). (Right) During the fall from the plateau, there is another increase in polarization, possibly due to the revealing of an asymmetric helium core (red oval) with a different elongation than any earlier structure (\autoref{subsec:img}).}
    \label{fig:sketch}
\end{figure*}

Initially, during the rising phase (+2.33 days), we detect a significant level of continuum polarization from the imaging polarimeter MOPTOP in $R$ band, $p_R=1.02 \pm 0.06 \%$ and  $PA_R=152.4  \pm 2.1 \degr$. This is consistent with the value of $p_R=0.82 \pm 0.07\%$ reported by \citet{Maund_2023} for the same epoch, and other values reported elsewhere during the rising phase \citep[]{Vasylyev_2023,Singh_2024_23ixfPol}. In this phase, the SN underwent significant CSM interaction, as seen via the flash features in the early spectra of SN~2023ixf \citep{Perlely_2023_23ixfClassification,Bostroem_2023_23ixf,Smith_2023,Jacobson-Galan_2023,Zimmerman_2024}. Under the assumption of electron scattering, this level of polarization points to a notable break in the spherical symmetry of the CSM, such as an elongated scattering geometry. Thus, we conclude that the CSM close to the progenitor is aspherical, which is also shown by high-resolution spectral data \citep{Smith_2023}. We also observe a high degree of intrinsic polarization in other filters during this epoch with $p_B = 0.99 \pm 0.07\%$, $p_V = 1.21 \pm 0.08\% $, $p_I = 0.85 \pm 0.09\%$ for $B$, $V$, and $I$ filters respectively. The $PA$ values for these filters are consistent with the $R$-band value with $PA_B = 158 \pm 4 \degr$, $PA_V = 155 \pm 4 \degr$, and $PA_I = 156 \pm 6 \degr$.

In the first epoch of SPOL data, at the start of the plateau phase (+12.40 days), we see some intrinsic continuum polarization, as also shown in \autoref{fig:qu_continuum} and \autoref{fig:magpol}. The $PA$ during this phase is $59 \pm 16 \degr$ which is close to a $90 \degr$ rotation compared to our observations during the rise phase (+2.33 days). This $90 \degr$ $PA$ change was seen by \citet{Vasylyev_2023} for 5--15 days after the explosion, as shown in \autoref{fig:magpol}.
This continuum polarization decreases over time, and by +17.41 days, its value goes to $0.16 \pm 0.07\%$). As we noted in Section 3.1, the continuum polarization reaches even lower levels during days 26-31.

The decline in continuum polarization indicates that the emission is not dominated by the CSM interaction, while the final values near $p=0.19\%$ suggest that from +17.41 to 31.57 days when the light curve is well settled in the plateau phase, the intrinsic polarization is very low. 
These polarization values and their evolution are consistent with those of the imaging polarimetry presented by \citet{Singh_2024_23ixfPol}. We note that the position angle gradually changes during this phase as shown in \autoref{fig:magpol} (bottom panel) with a larger error bar in $PA$ values. 
We caution that the polarization value during this time range is very low, thus the significance of the change in $PA$ is limited.

\begin{figure*}
    \centering
    \includegraphics[width=\textwidth]{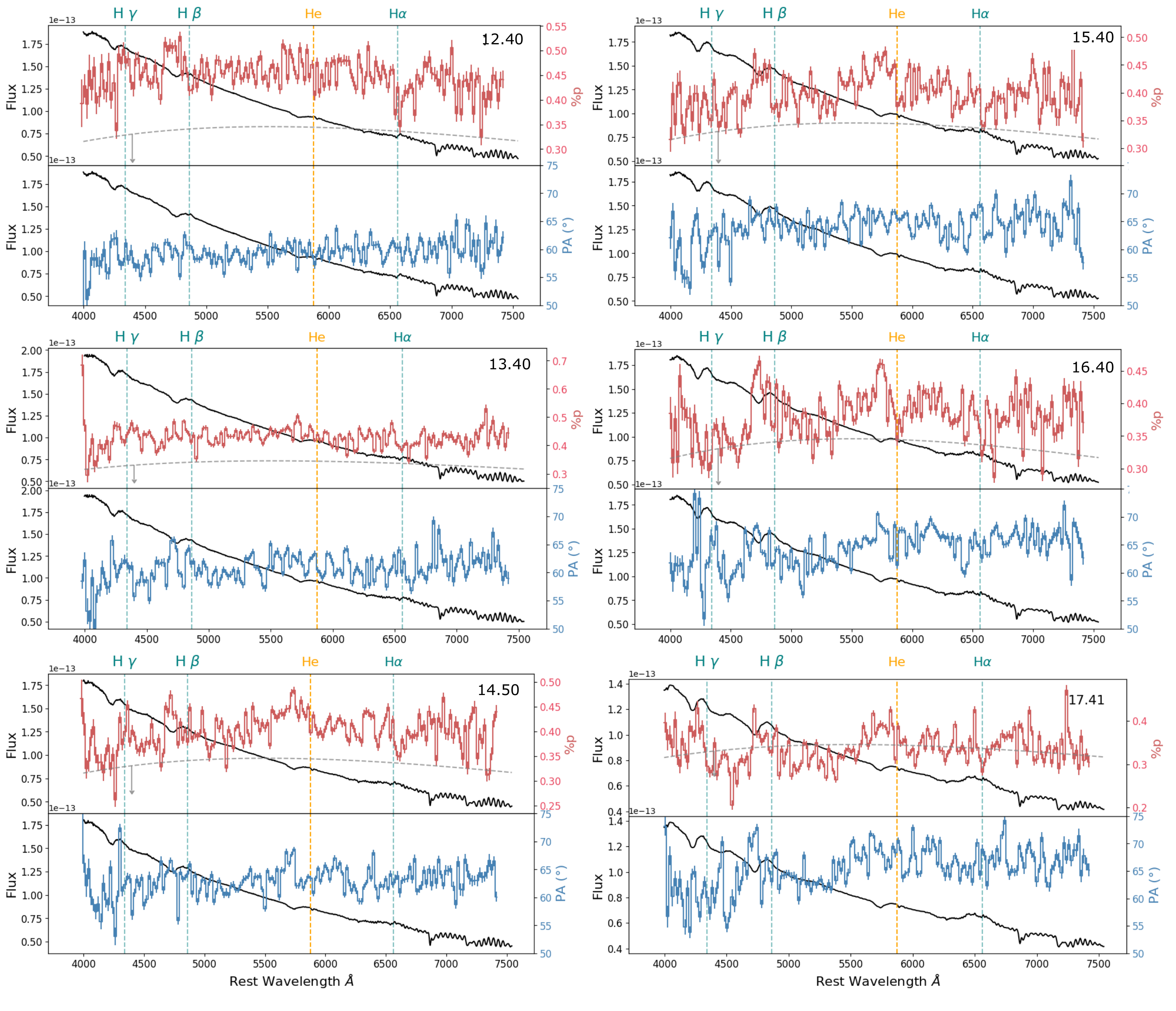}
    \caption{SPOL data from +12.40 days to +17.41 days after the explosion. For each epoch, the top panel contains relative flux (black) and $\%p$ (red) values and the bottom panel is for the $PA$ (blue), which is binned to 20 \AA. Major hydrogen and helium lines at rest wavelengths are overplotted. The light gray line in the $\%p$ panel is the maximum ISP estimate as discussed in \autoref{sec:ip}. There is a clear continuum polarization in the first two epochs and gradual evolution toward the ISP ($< 0.35\%$) with time. The $PA$ is fairly consistent for all the epochs presented here as shown in \autoref{tab:pol}. Finally, clear depolarizations are observed for some lines. The observed continuum polarization above the maximum ISP suggests some level of asphericity in the initial epoch of observation. }
    \label{fig:papol1}
\end{figure*}

By contrast, once the light curve falls from the plateau (+73.19 days; \citealt{hsu2024yearsn2023ixfbreaking}), the continuum polarization increases again to $0.45 \pm 0.08 \%$ (\autoref{fig:magpol}). We observe a similar polarization value for the subsequent epoch at +76.19 days ($0.62 \pm 0.08 \%$).
We note that we also detect significant polarization in other filters at this epoch: $p_B = 1.03 \pm 0.12 $, $p_V = 0.98 \pm 0.09 $, $p_I = 0.89 \pm 0.08 $ with position angles of $PA_B = 27 \pm 7 \degr$, $PA_V = 16 \pm 6 \degr$, and $PA_I = 16 \pm 5 \degr$for +73.19 days. For +76.19 days we find $p_B = 0.86 \pm 0.15 $, $p_V = 0.87 \pm 0.11 $, $p_I = 0.79 \pm 0.08 $, $PA_B = 172 \pm 10 \degr$, $PA_V = 177 \pm 8 \degr$, and $PA_I = 179 \pm 6 \degr$for $B$, $V$, and $I$ filters respectively. Additionally, we observe $PA_R = 22 \degr$ which is $\sim 130 \degr$ different from the first epoch of MOPTOP data. This indicates that the mechanism producing polarization during this phase is different from the initial phase.
Similar behavior in other type II SNe has been attributed to an aspherical core being revealed as the outer ejecta become transparent \citep[e.g.,][]{Leonard_2006, Nagao_2019,Nagao_2021,Nagao_2023}. However, recent studies \citep{Dessart_2011,Dessart_2024_polModel} have attributed this solely to a reduction in optical depth. 
Effectively, it is explained as a purely radiative transfer effect due to a reduction in polarimetric cancellation as the optical depth decreases and single scattering becomes dominant \citep{Dessart_2011,Dessart_2024_polModel}. We also observe some position angle rotation as the light curve falls from the plateau going from $166\degr \pm 25\degr$ (+31.57 days) to $22 \degr \pm 10\degr$ (+73.19 days). 

\cite{Singh_2024_23ixfPol} presented time-dependent $R$-band ISP corrected polarization of SN~2023ixf, identifying three different peaks at +1.4 days [($1.09\% \pm 0.05 \%$), ($153.4\degr\pm 0.3\degr$)], +6.4 days [($0.54\% \pm 0.06 \%$), ($60.3\degr\pm 1.1\degr$)], and +79.2 days [($0.48\% \pm 0.05 \%$), ($16.9\degr\pm 0.8\degr$)] days. We observe the first and last peaks in our data as well, however, we do not have an observation at +6.4 days. Their observed polarization amplitude and Stokes $q$ and $u$ evolution are consistent with our data during overlapping epochs. 

\autoref{fig:papol1} displays the polarization and $PA$ spectra we observed with SPOL between +12.40 and +17.41 days, with flux spectra and maximum ISP curves overplotted for comparison. 
At all these epochs, the continuum $PA$ is consistently within a range of $60\degr-75\degr$, which represents a rotation of almost $90\degr$ from our first observation during the rising phase ($152\degr$). This behavior is also apparent in \autoref{fig:qu_continuum}, where the first and second data points in phase are located in opposite quadrants. 
This significant rotation could imply that the photosphere has engulfed the CSM region during the peak in the light curve (+12.40 days) and there is a very low level of interaction happening in the CSM region (as suggested by the lack of narrow emission lines in the spectra) as it is swept away. We note that in their simulations,  \citet{Dessart_2024_spol_1998s} also found a drastic change in polarization direction at the conclusion of the strong interaction phase, and similarly attributed it to emission and scattering regions having different spatial distributions.

\citet{Morozova_2018} studied 20 Type II SNe and found that a significant fraction in their sample had light curve peaks that could not be explained by hydrodynamical models using SNEC without CSM interaction; instead the models with CSM provided a better fit, as shown in Figure~2 of \citet{Morozova_2018}. A similar excess in peak is seen for SN~2023ixf, which indicates that some CSM interaction is present at the time of the light curve peak and our first SPOL epoch ($\sim$+12 days). The emission is likely not dominated by the CSM interaction as we see during the rise phase, as there are no narrow emission lines during this phase. 
However, at this epoch, the presence of intrinsic continuum polarization and a flip in $PA$ by $90\degr$ from the earlier observation 
suggests that CSM interaction during the rise produced the ``pinched waist" geometry seen in strongly interacting SNe   \citep{Mauerhan_2014,Smith_2015}. Scattering by CSM in the pinched waist cancels some Stokes vectors, resulting in a $PA$ rotation of $90 \degr$.We note that CSM interaction during this phase has also been invoked for some type IIn and II SNe \citep[e.g.][]{Smith_2015,moriya_2012}.
As time passes, the interaction comes to an end and the photosphere is no longer constrained to an elongated shape. The light curve settles to a plateau phase, during which the CSM interaction does not play a major role. Most of the emission is from the photosphere, which retains its earlier preferred angle but approaches a spherical geometry, thus we do not detect significant intrinsic polarization during this phase \citep[see also][]{Vasylyev_2023}.  


\citet{Dessart_2024_polModel} performed 2D polarization radiative transfer calculations to simulate the behavior of the continuum polarization of type II SN from 20 to 300 days after the explosion. They used models following the prescription from their previous work on SN 2012aw \citep{Dessart_2021} but with greater kinetic energy and higher $^{56}$Ni abundances. In their models, they implemented various bipolar explosions in which the spherical symmetry was broken by adding different materials within $\sim 30 \degr$ of the poles. They calculated continuum polarization for the spectral range from 6900--7200 \AA. They found that the polarization peaks during the fall from the plateau, but the temporal evolution of polarization during the plateau phase is model-dependent. Observationally similar diversity in the evolution of polarization during the plateau phase has been seen in type II SN \citep{Nagao_2023}. 

Comparing the results from different models in \citet{Dessart_2024_polModel} with the behavior of our observed polarimetric data for  SN~2023ixf, we find that the model e1ni1b1/e1ni1 produces the closest evolution. More information about the different models can be found in Table 1 of \citet{Dessart_2024_polModel}.
The bipolar portion of this model, e1ni1b1, has a $^{56}$Ni core of 0.009 $M_\odot$ and a $^{56}$Ni shell of 0.02 $M_\odot$ that is distributed in the bipolar region within $30\degr$ from the pole. This is then paired with e1ni1 which makes up the rest of the scattering region with a $^{56}$Ni core of 0.009 $M_\odot$ and no $^{56}$Ni shell. We note that the $^{56}$Ni mass for SN~2023ixf from observations is higher than the values used in these models \citep{hsu2024yearsn2023ixfbreaking,Singh_2024_23ixfPol,Hiramatsu_2023,Moriya_2024}. From this model (Figure 8 left panel in \citealt{Dessart_2024_polModel}), the continuum polarization during the photospheric phase (once the light curve is fully settled to the plateau phase) is close to zero, consistent with SN~2023ixf. Then there is a jump in polarization during the fall from the plateau (Figure 8 left panel in \citealt{Dessart_2024_polModel}), similar to the observed data. This behavior was also seen in SN~2004dj \citep{Leonard_2006}. 
\citealt{Dessart_2024_polModel} attribute the increase in polarization at the end of the plateau phase to the transition from multiple to single scattering, increasing the polarization via reduced cancellation of Stokes parameters. This differs from other interpretations which attribute the change in polarization to an unveiling of the asymmetric inner helium ejecta or $^{56}$Ni bubbles/plumes during this transition phase \citep{Leonard_2001_typeII, Leonard_2006,Nagao_2021,Nagao_2023}. We note that we detect a significant change of $\sim20 \degr$ in $PA$ between +73.19 and +76.19 days, which could indicate that there is some inner structure producing this polarization that is not aligned to the scattering region during the plateau phase. We observe this change in position angle for all the filters.

\subsection{Line polarization}
\label{subsec:line}

\begin{figure}
    \centering
    \includegraphics[width=\columnwidth]{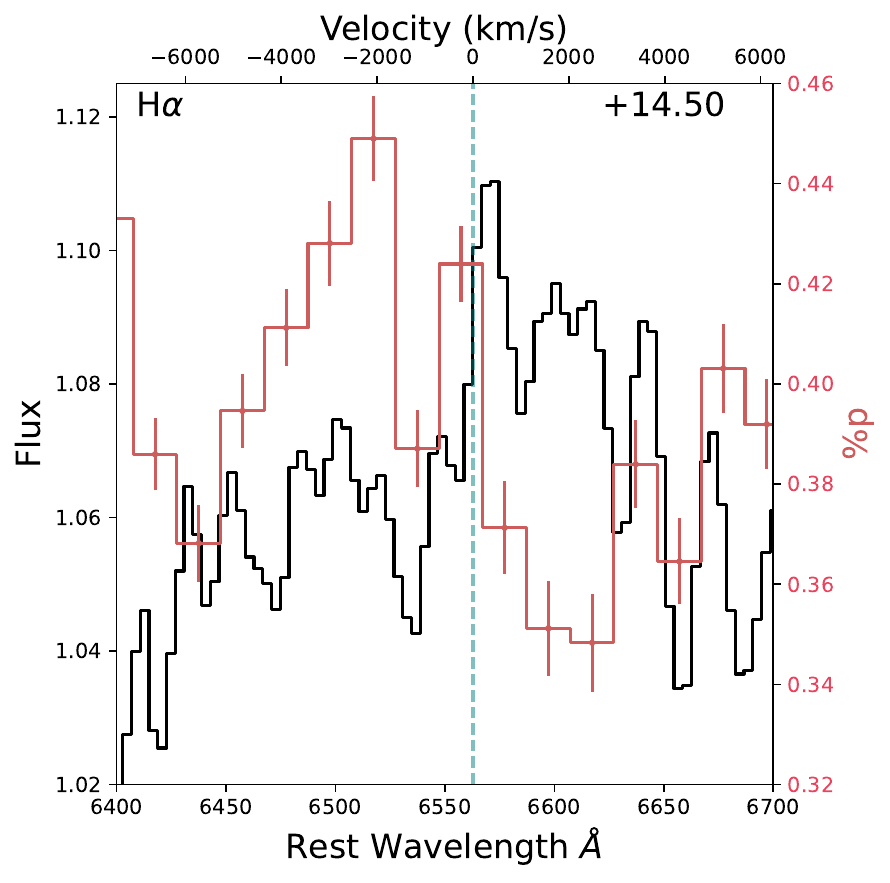}
    \caption{Relative flux (black) and polarization (red, binned to 20 \AA) of SN 2023ixf in the region near H$\alpha$ from our SPOL observation +14.50 days post-explosion. The comparison contrasts the P Cygni profile in the flux spectrum with the inverse P Cygni profile in polarization.}
    \label{fig:pcygni}
\end{figure}

The polarized spectra displayed in \autoref{fig:papol1} contain various interesting line polarization features. At day +12.40, we observe some depolarization at the H$\alpha$ ($\lambda 6563 $) and H$\gamma$ ($\lambda 4341$) lines, as shown by the fact that these lines agree in polarization with our ISP upper limit (less polarized than the continuum). We also see clear line polarization in the blue-shifted absorption of H$\beta$ ($\lambda 4861$), as marked in \autoref{fig:papol1}. 
At days +14.50 and +15.40 we see distinct inverse P Cygni profiles in polarization for H$\alpha$, H$\beta$, and He I ($\lambda 5876$); we present a zoomed-in figure for H$\alpha$ at this epoch in \autoref{fig:pcygni}. We observe small $PA$ rotations ($<5\degr$) associated with some of the inverse P Cygni profiles in SN 2023ixf, but these are on the same scale as the noise in the $PA$ spectra. We conclude that the position angles associated with these inverse P Cygni profiles during the plateau phase agree well with the overall continuum $PA$ ($60-70\degr$).

This inverse P Cygni polarization profile 
is a key indicator of asymmetry in SNe \citep{McCall_1984,Jeffery_1991_pcyg,Leonard_2001,Maund_2024}. The redshifted emission feature seen in the P Cygni flux profile is produced by resonant scattering of light into our line of sight by ions or atoms in the CSM. Because this resonant scattering polarizes light much less efficiently than electron scattering, it effectively dilutes the continuum polarization. This produces a polarization dip corresponding to the flux increase, as seen in \autoref{fig:pcygni}. On the other hand, the blue-shifted absorption feature seen in the P Cygni flux profile is associated with a peak in polarization, as the region producing this dip in flux blocks or redirects unpolarized light from the central source and thereby increases the fractional polarization. This phenomenon has been seen previously in two type II SNe, SN~1999em \citep{Leonard_2001} and SN~2021yja \citep{Vasylyev_2024}. Simply put, these inverse P-Cygni profiles are likely due to the intervening material associated with the remaining CSM, which both scatters unpolarized light into our sightline and obscures our view of the emission region. However, the lack of a strong $PA$ rotation across these lines suggests that the remaining CSM during the plateau phase is axisymmetrically distributed with respect to the ejecta \citep{Jeffery_1991_pcyg}. This is consistent with our picture of the remnant CSM at this stage being aligned with the ``waist" of the photosphere.

\section{Comparison with other SNe}\label{sec:comp}
\begin{figure}
    \centering
    \includegraphics[width=\columnwidth]{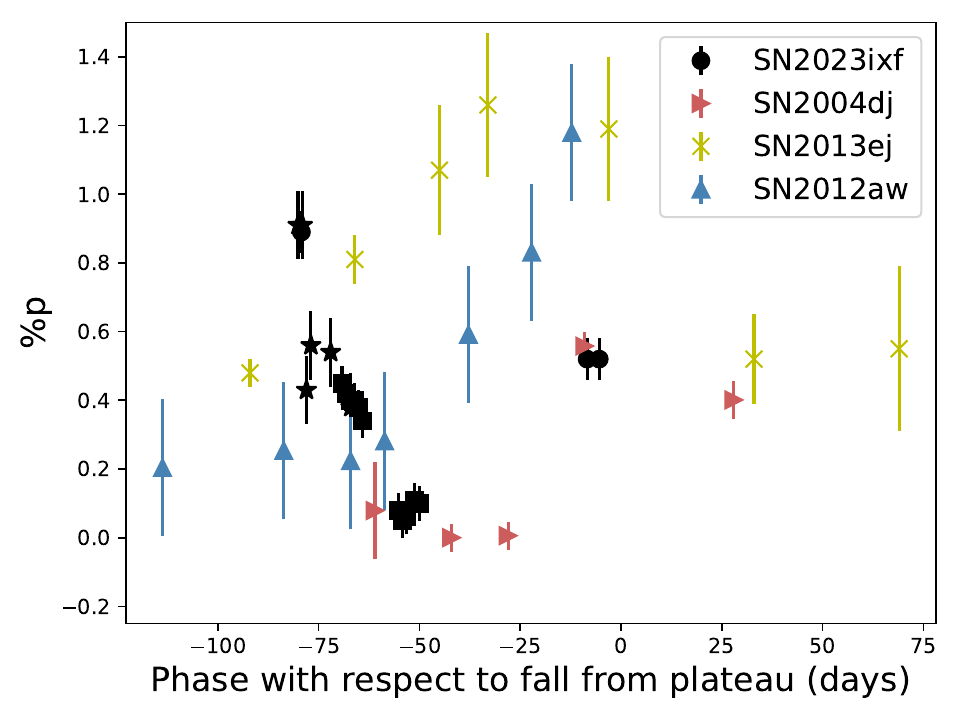}
    \caption{Evolution of the polarization level for various type II SNe as a function of phase with respect to the fall from plateau (+81.50 days). For SN~2023ixf, different symbols signify data from different instruments: MOPTOP (circle), SPOL (square), and Kast (star) \citep{Vasylyev_2023}. Red, blue, and yellow points refer to SN~2004dj, SN~2012aw, and SN~2013ej from \citet{Leonard_2006}, \citet{Dessart_2021}, and \citet{Nagao_2021} respectively. }
    \label{fig:pol_comp}
\end{figure}

\begin{figure}
    \centering
    \includegraphics[width=\columnwidth]{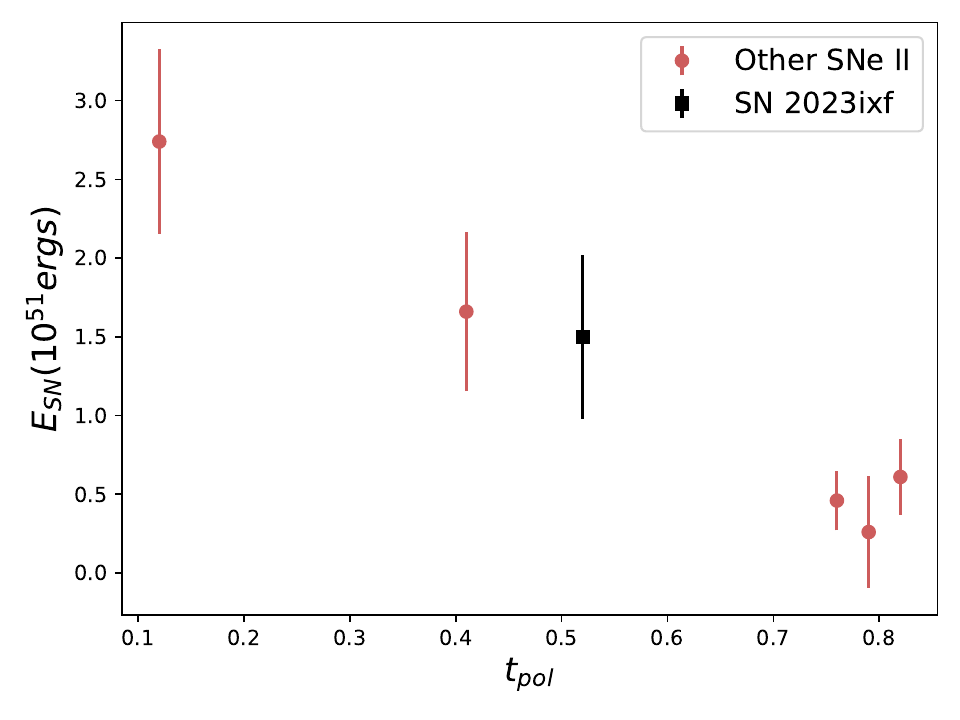}
    \caption{Explosion energy of SNe 2012aw, 2017gmr, 2007aa, 2006ov, and 2004dj from \citet{Nagao_2023} (red) along with SN~2023ixf (black) with respect to $t_{pol}$ (as defined in \autoref{sec:comp}). For SN~2023ixf the explosion energy has been estimated to be 1 to 2 $\times 10^{51}$ ergs \citep{Singh_2024_23ixfPol, Hiramatsu_2023}, hence we use the center of this range as our explosion energy value and the error bar spans these two values. The calculated value of SN~2023ixf follows the relation seen for other Type II SNe in the literature.}
    \label{fig:esn_comp}
\end{figure}

We compare the polarization behavior of SN~2023ixf with other Type II SNe in \autoref{fig:pol_comp} and \autoref{fig:esn_comp}. In \autoref{fig:pol_comp}, we compare the polarization evolution of SN~2023ixf from this work as well as values reported in \citet{Vasylyev_2023} and \citet{Singh_2024_23ixfPol} along with SN~2004dj \citep{Leonard_2006}, SN~2012aw \citep{Dessart_2021}, and SN~2013ej \citep{Nagao_2021}. In this figure, the phases are days since the fall from the plateau or $T_{pt}$ as described in \citet{Valenti_2016}. We chose to present these three because they show a rise in polarization when they fall from the plateau. For both SN~2004dj and SN~2012aw, there are no polarization observations during the very early phase when the brightness is increasing. Thus, we cannot assess whether these SNe showed an initial polarization peak similar to SN~2023ixf. However, for SN~2013ej \citet{Nagao_2021} contribute the early polarization detection to CSM interaction.

During the plateau phase, the observed polarization behavior of SN~2023ixf and SN~2004dj is similar and very close to no intrinsic polarization. However, SN~2012aw and SN~2013ej behave differently with polarization increasing with time and reaching the peak when the light curve falls from the plateau. We note that we do not have data between +31.57 days and +73.19 days, hence the behavior of SN~2023ixf could be similar to SN~2012aw and SN~2013ej. 

\citet{Nagao_2023} examined 15 different Type II SNe and grouped them into 3 different categories based on polarization values during the photospheric phase. \citet{Nagao_2023} compares various properties of the SNe in their sample with the characteristic property of the continuum polarization which is calculated by normalizing the time of the polarization rise to the length of the photospheric phase ($t_\mathrm{pol} = T(\mathrm{polarization~rise})/T(\mathrm{photospheric~phase}) = 0.52$). \citet{Nagao_2023} found a clear correlation between the timing of the polarization rise and the explosion energy, indicating that the explosion asphericity is proportional to the explosion energy. SN~2023ixf falls into the Group 1 category comprised of SNe which also show a low polarization level at the photospheric phase and an increase in the value during the transitional phase from the photospheric to the tail. According to these criteria, SNe 2008bk, 2007aa, 2006ov, and 2004dj fall in the same category in their sample. However, SN~2012aw falls in Group 2 in \cite{Nagao_2023}, where the polarization increases during the photospheric phase, indicating that the SNe has an aspherical structure in general and not just the helium core. We have presented the relation of explosion energy of SN with respect to $t_{pol}$ in \autoref{fig:esn_comp}, where the red points are from the sample of \citet{Nagao_2023} and the black point is SN~2023ixf. We use an explosion energy of $1.5 \times 10^{51}$ erg, the midpoint of the range from \citet{Singh_2024_23ixfPol}, for SN~2023ixf and find behavior consistent with respect to the sample.

 Overall, SN~2023ixf behaves similarly to other hydrogen-rich SNe as seen in \autoref{fig:esn_comp}. However, due to its proximity and early follow-up, we detect a high level of polarization in the very early phase where the CSM interaction is dominant; such a signature has not been seen before for other normal type II SNe. The observed early polarization behavior of SN~2023ixf is reminiscent of Type IIn SNe which have longer-lasting CSM interaction. The polarization of Type IIn is relatively high and persistent but eventually drops at later times \citep[e.g.,][]{Hoffman_2008_97eg,Bilinski_2023}.

\section{Discussion \& Conclusions} \label{sec:conclusions}

In this paper, we present $R$-band imaging polarimetry and spectropolarimetry observations (and MOPTOP $B$, $V$, and $I$ data in \autoref{subsec:img}) of the nearby Type II SN~2023ixf. Our first imaging polarimetric data were taken +2.33 days after the explosion when flash features are present in the spectra and the light curve is rising to the peak, and the last data point was taken around the time when the photospheric phase ends and the fall from the plateau phase begins, i.e., +76.19 days. The first spectropolarimetric observation was done on +12.40 days and the last was taken on +31.57 days after the explosion. These data provide us with the following main takeaways:
\begin{enumerate}
    \item During the initial phase (+2.33 days) we detect an intrinsic polarization value of $1.02\% \pm 0.12\%$ and a $PA$ of 152\degr $\pm$ 14\degr.
    At this phase, there is also strong evidence of CSM interaction from spectroscopic and photometric data. Our detection of intrinsic polarization at this time confirms that the CSM is aspherical as suggested earlier by \citet{Smith_2023, Vasylyev_2023,Singh_2024_23ixfPol,Ferrari_2024_23ixf_nebular,Fang_2024_23ixfNebular}.
    \item Our next data set starts at +12.40 days, which is close to the maximum in the light curve going to the photospheric phase. The intrinsic continuum polarization is ($\sim$0.26$ \pm 0.07\%$) and there is a rotation in $PA$ by $90\degr$ compared to the +2.33 phase. This could indicate that as the photosphere expands during the rise phase, it is constrained by the CSM and becomes elongated perpendicular to the initial axis in a ``pinched waist'' geometry.
     After the peak at +12.40 days, the photosphere gradually engulfs the CSM and becomes more and more spherical, causing the continuum polarization to decrease toward the ISP value as the CSM is swept away. In addition, during the plateau phase, we see inverse P Cygni behavior in the H and He lines, and the $PA$ around these lines is consistent with the continuum $PA$. These inverse P Cygni profiles can be attributed to intervening remnant material that is obscuring our view from the emission region. 
    \item At day +73.19 of our observations, when the light curve falls from the plateau, we measure an increase in polarization to $0.45\%\pm 0.08\%$ and an increase in $PA$ to $22 \pm 10 \degr$. A similar jump in polarization during this phase has also been seen for other type II SNe in the literature. This jump can be explained in two different ways. First, the H envelope becomes optically thin and we see an aspherical He core. Alternatively, the optical depth of the scattering medium decreases to the point where single scattering is dominant, which in turn increases the polarization. We have a change in position angle during the fall from the plateau, and this could indicate that we are observing a scattering medium (possibly inner He core) that is not in the same orientation as the scattering medium producing light during the plateau phase. 
    \item We also compare SN~2023ixf imaging polarimetry behavior in the $R$ band with other Type II SNe. We find that from the photospheric phase to the plateau falloff phase, SN~2023ixf behaves similarly to other type II SNe from the literature. However, the early high level of polarization during the rise phase has not been seen before for Type II SNe that are not Type IIn.

\end{enumerate}

To encapsulate the information gathered from our multi-epoch polarimetry data set, we present a simplified sketch of the temporal evolution of SN~2023ixf in \autoref{fig:sketch}. The initial polarization peak can be attributed to an aspherical CSM where electron scattering produces a polarization signal perpendicular to the scattering medium. This marks the first detection of high-level polarization at such an early time for any Type II SNe to date. As the photosphere expands, the $PA$ rotates by $90\degr$ between the light-curve rise and the light-curve plateau. Apart from the early high level of polarization, we find the general correlation between explosion energy and $t_{pol}$ seen in other observational studies  \citep[e.g.][]{Nagao_2023} holds for SN~2023ixf. In particular, the behavior from the plateau phase to the fall from the plateau is similar to that previously seen in SN~2004dj. Our work shows that polarimetric observations can provide complementary information to spectroscopic and photometric observations about the SN, its mass loss history, and its explosion mechanism. Additionally, we show that early polarimetry observations give additional information about the mass loss from massive stars during their final stages of evolution before the explosion. Together, these findings motivate the need for more rapid polarimetry follow-up observations of type II SN. 


\section{acknowledgments}
We would like to thank the anonymous referee for constructive comments which have improved the paper immensely. Time-domain research by the University of Arizona team, M.S. and D.J.S.\ is supported by National Science Foundation (NSF) grants 2108032, 2308181, 2407566, and 2432036 and the Heising-Simons Foundation under grant \#2020-1864.  Research by Y.D., S.V., N.M.R, D.M. and E.H. is supported by NSF grant AST-2008108. Operation of LT on the island of La Palma by Liverpool John Moores University at the Spanish Observatorio del Roque de los Muchachos of the Instituto de Astrofisica de Canarias is financially supported by the UK Science and Technologies Facilities Council (STFC). K.A.B. is supported by an LSSTC Catalyst Fellowship; this publication was thus made possible through the support of Grant 62192 from the John Templeton Foundation to LSSTC. The opinions expressed in this publication are those of the authors and do not necessarily reflect the views of LSSTC or the John Templeton Foundation. JEA is supported by the international Gemini Observatory, a program of NSF's NOIRLab, which is managed by the Association of Universities for Research in Astronomy (AURA) under a cooperative agreement with the National Science Foundation, on behalf of the Gemini partnership of Argentina, Brazil, Canada, Chile, the Republic of Korea, and the United States of America. JAC-B acknowledges support from FONDECYT Regular N 1220083. SD and JLH  acknowledge support from NSF award AST-2009996. They also recognize that the University of Denver resides on the ancestral territories of the Arapaho, Cheyenne, and Ute nations and that its history is inextricably linked with the violent displacement of these indigenous peoples. D.C.L. acknowledges support from NSF grant  AST-2010001.
This work makes use of data taken with the Las Cumbres Observatory global telescope network. The LCO group is supported by NSF grants 1911225 and 1911151. 
This research has made use of the NASA Astrophysics Data System (ADS) Bibliographic Services, and the NASA/IPAC Infrared Science Archive (IRSA), which is funded by the National Aeronautics and Space Administration and operated by the California Institute of Technology.  This research made use of Photutils, an Astropy package for detection and photometry of astronomical sources (\cite{Bradley_2019}). This work made use of data supplied by the UK Swift Science Data Centre at the University of Leicester.
Observations reported here were obtained at the MMT Observatory, a joint facility of the University of Arizona and the Smithsonian Institution.  
This research has made use of the CfA Supernova Archive, which is funded in part by the National Science Foundation through grant AST 0907903.


%

\vspace{5mm}
\facilities{Bok (SPOL), Liverpool Telescope (MOPTOP)}


\software{Astropy \citep{astropy:2013,astropy:2018, astropy:2022}, Photutils \citep{Bradley_2019},  Matplotlib \citep{mpl}, Numpy \citep{numpy}, Scipy \citep{scipy}, IRAF \citep{iraf1,iraf2}, \texttt{lcogtsnpipe}\citep{Valenti_2016}
          }

\bibliography{SN23ixf_pol}{}

\newcommand{\noop}[1]{}
\begin{thebibliography}{}
\expandafter\ifx\csname natexlab\endcsname\relax\def\natexlab#1{#1}\fi
\providecommand{\url}[1]{\href{#1}{#1}}
\providecommand{\dodoi}[1]{doi:~\href{http://doi.org/#1}{\nolinkurl{#1}}}
\providecommand{\doeprint}[1]{\href{http://ascl.net/#1}{\nolinkurl{http://ascl.net/#1}}}
\providecommand{\doarXiv}[1]{\href{https://arxiv.org/abs/#1}{\nolinkurl{https://arxiv.org/abs/#1}}}

\bibitem[{{Astropy Collaboration} {et~al.}(2013){Astropy Collaboration}, {Robitaille}, {Tollerud}, {Greenfield}, {Droettboom}, {Bray}, {Aldcroft}, {Davis}, {Ginsburg}, {Price-Whelan}, {Kerzendorf}, {Conley}, {Crighton}, {Barbary}, {Muna}, {Ferguson}, {Grollier}, {Parikh}, {Nair}, {Unther}, {Deil}, {Woillez}, {Conseil}, {Kramer}, {Turner}, {Singer}, {Fox}, {Weaver}, {Zabalza}, {Edwards}, {Azalee Bostroem}, {Burke}, {Casey}, {Crawford}, {Dencheva}, {Ely}, {Jenness}, {Labrie}, {Lim}, {Pierfederici}, {Pontzen}, {Ptak}, {Refsdal}, {Servillat}, \& {Streicher}}]{astropy:2013}
{Astropy Collaboration}, {Robitaille}, T.~P., {Tollerud}, E.~J., {et~al.} 2013, \aap, 558, A33, \dodoi{10.1051/0004-6361/201322068}

\bibitem[{{Astropy Collaboration} {et~al.}(2022){Astropy Collaboration}, {Price-Whelan}, {Lim}, {Earl}, {Starkman}, {Bradley}, {Shupe}, {Patil}, {Corrales}, {Brasseur}, {N{\"o}the}, {Donath}, {Tollerud}, {Morris}, {Ginsburg}, {Vaher}, {Weaver}, {Tocknell}, {Jamieson}, {van Kerkwijk}, {Robitaille}, {Merry}, {Bachetti}, {G{\"u}nther}, {Aldcroft}, {Alvarado-Montes}, {Archibald}, {B{\'o}di}, {Bapat}, {Barentsen}, {Baz{\'a}n}, {Biswas}, {Boquien}, {Burke}, {Cara}, {Cara}, {Conroy}, {Conseil}, {Craig}, {Cross}, {Cruz}, {D'Eugenio}, {Dencheva}, {Devillepoix}, {Dietrich}, {Eigenbrot}, {Erben}, {Ferreira}, {Foreman-Mackey}, {Fox}, {Freij}, {Garg}, {Geda}, {Glattly}, {Gondhalekar}, {Gordon}, {Grant}, {Greenfield}, {Groener}, {Guest}, {Gurovich}, {Handberg}, {Hart}, {Hatfield-Dodds}, {Homeier}, {Hosseinzadeh}, {Jenness}, {Jones}, {Joseph}, {Kalmbach}, {Karamehmetoglu}, {Ka{\l}uszy{\'n}ski}, {Kelley}, {Kern}, {Kerzendorf}, {Koch}, {Kulumani}, {Lee}, {Ly}, {Ma}, {MacBride}, {Maljaars}, {Muna}, {Murphy}, {Norman},
  {O'Steen}, {Oman}, {Pacifici}, {Pascual}, {Pascual-Granado}, {Patil}, {Perren}, {Pickering}, {Rastogi}, {Roulston}, {Ryan}, {Rykoff}, {Sabater}, {Sakurikar}, {Salgado}, {Sanghi}, {Saunders}, {Savchenko}, {Schwardt}, {Seifert-Eckert}, {Shih}, {Jain}, {Shukla}, {Sick}, {Simpson}, {Singanamalla}, {Singer}, {Singhal}, {Sinha}, {Sip{\H{o}}cz}, {Spitler}, {Stansby}, {Streicher}, {{\v{S}}umak}, {Swinbank}, {Taranu}, {Tewary}, {Tremblay}, {de Val-Borro}, {Van Kooten}, {Vasovi{\'c}}, {Verma}, {de Miranda Cardoso}, {Williams}, {Wilson}, {Winkel}, {Wood-Vasey}, {Xue}, {Yoachim}, {Zhang}, {Zonca}, \& {Astropy Project Contributors}}]{astropy:2022}
{Astropy Collaboration}, {Price-Whelan}, A.~M., {Lim}, P.~L., {et~al.} 2022, \apj, 935, 167, \dodoi{10.3847/1538-4357/ac7c74}

\bibitem[{{Bilinski} {et~al.}(2023){Bilinski}, {Smith}, {Williams}, {Smith}, {Leonard}, {Hoffman}, {Andrews}, \& {Milne}}]{Bilinski_2023}
{Bilinski}, C., {Smith}, N., {Williams}, G.~G., {et~al.} 2023, arXiv e-prints, arXiv:2304.13034, \dodoi{10.48550/arXiv.2304.13034}

\bibitem[{{Bostroem} {et~al.}(2023){Bostroem}, {Pearson}, {Shrestha}, {Sand}, {Valenti}, {Jha}, {Andrews}, {Smith}, {Terreran}, {Green}, {Dong}, {Lundquist}, {Haislip}, {Hoang}, {Hosseinzadeh}, {Janzen}, {Jencson}, {Kouprianov}, {Paraskeva}, {Meza Retamal}, {Reichart}, {Arcavi}, {Bonanos}, {Coughlin}, {Farah}, {Hawley}, {Hebb}, {Hiramatsu}, {Howell}, {Iijima}, {Ilyin}, {McCully}, {Moran}, {Morris}, {Mura}, {Newsome}, {Pabst}, {Ochner}, {Padilla Gonzalez}, {Pastorello}, {Pellegrino}, {Ravi}, {Reguitti}, {Salo}, {Vinko}, {Wheeler}, {Williams}, \& {Wyatt}}]{Bostroem_2023_23ixf}
{Bostroem}, K.~A., {Pearson}, J., {Shrestha}, M., {et~al.} 2023, arXiv e-prints, arXiv:2306.10119, \dodoi{10.48550/arXiv.2306.10119}

\bibitem[{Bradley {et~al.}(2019)Bradley, Sip{\H o}cz, Robitaille, Tollerud, Vin{\'{\i}}cius, Deil, Barbary, G{\"u}nther, Cara, Busko, Conseil, Droettboom, Bostroem, Bray, Bratholm, Wilson, Craig, Barentsen, Pascual, Donath, Greco, Perren, Lim, \& Kerzendorf}]{Bradley_2019}
Bradley, L., Sip{\H o}cz, B., Robitaille, T., {et~al.} 2019, astropy/photutils: v0.6, \dodoi{10.5281/zenodo.2533376}

\bibitem[{{Brown} {et~al.}(2013){Brown}, {Baliber}, {Bianco}, {Bowman}, {Burleson}, {Conway}, {Crellin}, {Depagne}, {De Vera}, {Dilday}, {Dragomir}, {Dubberley}, {Eastman}, {Elphick}, {Falarski}, {Foale}, {Ford}, {Fulton}, {Garza}, {Gomez}, {Graham}, {Greene}, {Haldeman}, {Hawkins}, {Haworth}, {Haynes}, {Hidas}, {Hjelstrom}, {Howell}, {Hygelund}, {Lister}, {Lobdill}, {Martinez}, {Mullins}, {Norbury}, {Parrent}, {Paulson}, {Petry}, {Pickles}, {Posner}, {Rosing}, {Ross}, {Sand}, {Saunders}, {Shobbrook}, {Shporer}, {Street}, {Thomas}, {Tsapras}, {Tufts}, {Valenti}, {Vander Horst}, {Walker}, {White}, \& {Willis}}]{Brown_2013}
{Brown}, T.~M., {Baliber}, N., {Bianco}, F.~B., {et~al.} 2013, \pasp, 125, 1031, \dodoi{10.1086/673168}

\bibitem[{{Bruch} {et~al.}(2021){Bruch}, {Gal-Yam}, {Schulze}, {Yaron}, {Yang}, {Soumagnac}, {Rigault}, {Strotjohann}, {Ofek}, {Sollerman}, {Masci}, {Barbarino}, {Ho}, {Fremling}, {Perley}, {Nordin}, {Cenko}, {Adams}, {Adreoni}, {Bellm}, {Blagorodnova}, {Bulla}, {Burdge}, {De}, {Dhawan}, {Drake}, {Duev}, {Dugas}, {Graham}, {Graham}, {Irani}, {Jencson}, {Karamehmetoglu}, {Kasliwal}, {Kim}, {Kulkarni}, {Kupfer}, {Liang}, {Mahabal}, {Miller}, {Prince}, {Riddle}, {Sharma}, {Smith}, {Taddia}, {Taggart}, {Walters}, \& {Yan}}]{Bruch_2021}
{Bruch}, R.~J., {Gal-Yam}, A., {Schulze}, S., {et~al.} 2021, \apj, 912, 46, \dodoi{10.3847/1538-4357/abef05}

\bibitem[{{Cardelli} {et~al.}(1989){Cardelli}, {Clayton}, \& {Mathis}}]{Cardelli_1989}
{Cardelli}, J.~A., {Clayton}, G.~C., \& {Mathis}, J.~S. 1989, \apj, 345, 245, \dodoi{10.1086/167900}

\bibitem[{{Chornock} {et~al.}(2010){Chornock}, {Filippenko}, {Li}, \& {Silverman}}]{Chornock_2010}
{Chornock}, R., {Filippenko}, A.~V., {Li}, W., \& {Silverman}, J.~M. 2010, \apj, 713, 1363, \dodoi{10.1088/0004-637X/713/2/1363}

\bibitem[{{Dessart} \& {Hillier}(2011)}]{Dessart_2011}
{Dessart}, L., \& {Hillier}, D.~J. 2011, \mnras, 415, 3497, \dodoi{10.1111/j.1365-2966.2011.18967.x}

\bibitem[{{Dessart} {et~al.}(2024{\natexlab{a}}){Dessart}, {Hillier}, \& {Leonard}}]{Dessart_2024_polModel}
{Dessart}, L., {Hillier}, D.~J., \& {Leonard}, D.~C. 2024{\natexlab{a}}, \aap, 684, A16, \dodoi{10.1051/0004-6361/202347808}

\bibitem[{{Dessart} {et~al.}(2021){Dessart}, {Leonard}, {Hillier}, \& {Pignata}}]{Dessart_2021}
{Dessart}, L., {Leonard}, D.~C., {Hillier}, D.~J., \& {Pignata}, G. 2021, \aap, 651, A19, \dodoi{10.1051/0004-6361/202140281}

\bibitem[{{Dessart} {et~al.}(2024{\natexlab{b}}){Dessart}, {Leonard}, {Vasylyev}, \& {Hillier}}]{Dessart_2024_spol_1998s}
{Dessart}, L., {Leonard}, D.~C., {Vasylyev}, S.~S., \& {Hillier}, D.~J. 2024{\natexlab{b}}, arXiv e-prints, arXiv:2409.13562, \dodoi{10.48550/arXiv.2409.13562}

\bibitem[{{Fang} {et~al.}(2024){Fang}, {Moriya}, {Ferrari}, {Maeda}, {Folatelli}, {Ertini}, {Kuncarayakti}, {Andrews}, \& {Matsumoto}}]{Fang_2024_23ixfNebular}
{Fang}, Q., {Moriya}, T.~J., {Ferrari}, L., {et~al.} 2024, arXiv e-prints, arXiv:2409.03540, \dodoi{10.48550/arXiv.2409.03540}

\bibitem[{{Ferrari} {et~al.}(2024){Ferrari}, {Folatelli}, {Ertini}, {Kuncarayakti}, \& {Andrews}}]{Ferrari_2024_23ixf_nebular}
{Ferrari}, L., {Folatelli}, G., {Ertini}, K., {Kuncarayakti}, H., \& {Andrews}, J.~E. 2024, \aap, 687, L20, \dodoi{10.1051/0004-6361/202450440}

\bibitem[{{Gal-Yam} {et~al.}(2014){Gal-Yam}, {Arcavi}, {Ofek}, {Ben-Ami}, {Cenko}, {Kasliwal}, {Cao}, {Yaron}, {Tal}, {Silverman}, {Horesh}, {De Cia}, {Taddia}, {Sollerman}, {Perley}, {Vreeswijk}, {Kulkarni}, {Nugent}, {Filippenko}, \& {Wheeler}}]{GalYam_2014}
{Gal-Yam}, A., {Arcavi}, I., {Ofek}, E.~O., {et~al.} 2014, \nat, 509, 471, \dodoi{10.1038/nature13304}

\bibitem[{Harris {et~al.}(2020)Harris, Millman, van~der Walt, Gommers, Virtanen, Cournapeau, Wieser, Taylor, Berg, Smith, Kern, Picus, Hoyer, van Kerkwijk, Brett, Haldane, del R{\'{i}}o, Wiebe, Peterson, G{\'{e}}rard-Marchant, Sheppard, Reddy, Weckesser, Abbasi, Gohlke, \& Oliphant}]{numpy}
Harris, C.~R., Millman, K.~J., van~der Walt, S.~J., {et~al.} 2020, Nature, 585, 357, \dodoi{10.1038/s41586-020-2649-2}

\bibitem[{{Hiramatsu} {et~al.}(2023){Hiramatsu}, {Tsuna}, {Berger}, {Itagaki}, {Goldberg}, {Gomez}, {Kishalay De}, {Hosseinzadeh}, {Bostroem}, {Brown}, {Arcavi}, {Bieryla}, {Blanchard}, {Esquerdo}, {Farah}, {Howell}, {Matsumoto}, {McCully}, {Newsome}, {Gonzalez}, {Pellegrino}, {Rhee}, {Terreran}, {Vink{\'o}}, \& {Wheeler}}]{Hiramatsu_2023}
{Hiramatsu}, D., {Tsuna}, D., {Berger}, E., {et~al.} 2023, \apjl, 955, L8, \dodoi{10.3847/2041-8213/acf299}

\bibitem[{{Hoffman} {et~al.}(2008){Hoffman}, {Leonard}, {Chornock}, {Filippenko}, {Barth}, \& {Matheson}}]{Hoffman_2008_97eg}
{Hoffman}, J.~L., {Leonard}, D.~C., {Chornock}, R., {et~al.} 2008, \apj, 688, 1186, \dodoi{10.1086/592261}

\bibitem[{{Hosseinzadeh} {et~al.}(2023){Hosseinzadeh}, {Farah}, {Shrestha}, {Sand}, {Dong}, {Brown}, {Bostroem}, {Valenti}, {Jha}, {Andrews}, {Arcavi}, {Haislip}, {Hiramatsu}, {Hoang}, {Howell}, {Janzen}, {Jencson}, {Kouprianov}, {Lundquist}, {McCully}, {Meza Retamal}, {Modjaz}, {Newsome}, {Padilla Gonzalez}, {Pearson}, {Pellegrino}, {Ravi}, {Reichart}, {Smith}, {Terreran}, \& {Vink{\'o}}}]{Hosseinzadeh_2023_23ixf}
{Hosseinzadeh}, G., {Farah}, J., {Shrestha}, M., {et~al.} 2023, arXiv e-prints, arXiv:2306.06097, \dodoi{10.48550/arXiv.2306.06097}

\bibitem[{{Howell} \& {Global Supernova Project}(2017)}]{Howell_2017}
{Howell}, D.~A., \& {Global Supernova Project}. 2017, in American Astronomical Society Meeting Abstracts, Vol. 230, American Astronomical Society Meeting Abstracts \#230, 318.03

\bibitem[{Hsu {et~al.}(2024)Hsu, Smith, Goldberg, Bostroem, Hosseinzadeh, Sand, Pearson, Hiramatsu, Andrews, Beasor, Dong, Farah, Galbany, Gomez, Gonzalez, Gutiérrez, Howell, Könyves-Tóth, McCully, Newsome, Shrestha, Terreran, Villar, \& Wang}]{hsu2024yearsn2023ixfbreaking}
Hsu, B., Smith, N., Goldberg, J.~A., {et~al.} 2024.
\newblock \doarXiv{2408.07874}

\bibitem[{{Hunter}(2007)}]{mpl}
{Hunter}, J.~D. 2007, Computing in Science and Engineering, 9, 90, \dodoi{10.1109/MCSE.2007.55}

\bibitem[{{Itagaki}(2023)}]{Itagaki_2023_23ixf_discovery}
{Itagaki}, K. 2023, Transient Name Server Discovery Report, 2023-1158, 1

\bibitem[{{Jacobson-Gal{\'a}n} {et~al.}(2023){Jacobson-Gal{\'a}n}, {Dessart}, {Margutti}, {Chornock}, {Foley}, {Kilpatrick}, {Jones}, {Taggart}, {Angus}, {Bhattacharjee}, {Braff}, {Brethauer}, {Burgasser}, {Cao}, {Carlile}, {Chambers}, {Coulter}, {Dominguez-Ruiz}, {Dickinson}, {de Boer}, {Gagliano}, {Gall}, {Gao}, {Gates}, {Gomez}, {Guolo}, {Halford}, {Hjorth}, {Huber}, {Johnson}, {Karpoor}, {Laskar}, {LeBaron}, {Li}, {Lin}, {Loch}, {Lynam}, {Magnier}, {Maloney}, {Matthews}, {McDonald}, {Miao}, {Milisavljevic}, {Pan}, {Pradyumna}, {Ransome}, {Rees}, {Rest}, {Rojas-Bravo}, {Sandford}, {Ascencio}, {Sanjaripour}, {Savino}, {Sears}, {Sharei}, {Smartt}, {Softich}, {Theissen}, {Tinyanont}, {Tohfa}, {Villar}, {Wang}, {Wainscoat}, {Westerling}, {Wiston}, {Wozniak}, {Yadavalli}, \& {Zenati}}]{Jacobson-Galan_2023}
{Jacobson-Gal{\'a}n}, W.~V., {Dessart}, L., {Margutti}, R., {et~al.} 2023, \apjl, 954, L42, \dodoi{10.3847/2041-8213/acf2ec}

\bibitem[{{Jeffery}(1991)}]{Jeffery_1991_pcyg}
{Jeffery}, D.~J. 1991, \apj, 375, 264, \dodoi{10.1086/170187}

\bibitem[{{Jermak} {et~al.}(2016){Jermak}, {Steele}, \& {Smith}}]{Jermak_2016}
{Jermak}, H., {Steele}, I.~A., \& {Smith}, R.~J. 2016, in Society of Photo-Optical Instrumentation Engineers (SPIE) Conference Series, Vol. 9908, \procspie, 99084I, \dodoi{10.1117/12.2232324}

\bibitem[{{Jermak} {et~al.}(2018){Jermak}, {Steele}, \& {Smith}}]{Jermak_2018}
{Jermak}, H., {Steele}, I.~A., \& {Smith}, R.~J. 2018, in Society of Photo-Optical Instrumentation Engineers (SPIE) Conference Series, Vol. 10702, \procspie, 107024Q, \dodoi{10.1117/12.2312132}

\bibitem[{{Jones} {et~al.}(2023){Jones}, {French}, {Agnello}, {Angus}, {Ansari}, {Arendse}, {Gall}, {Grillo}, {Bruun}, {Hede}, {Hjorth}, {Izzo}, {Korhonen}, {Raimundo}, {Ramanah}, {Sarangi}, {Wojtak}, {Pfister}, {Auchettl}, {Chambers}, {Huber}, {Magnier}, {Boer}, {Fairlamb}, {Lin}, {Wainscoat}, {Lowe}, {Gao}, {Bulger}, {Schultz}, {Minguez}, {Engel}, {Gagliano}, {Narayan}, {Soraisam}, {Wang}, {Rest}, {Smartt}, {Smith}, {Alexander}, {Blanchard}, {DeMarchi}, {Hajela}, {Jacobson-Galan}, {Margutti}, {Matthews}, {Stauffer}, {Stroh}, {Terreran}, {Drout}, {Coulter}, {Dimitriadis}, {Foley}, {Hung}, {Kilpatrick}, {Rojas-Bravo}, {Siebert}, \& {Ramirez-Ruiz}}]{Jones_2023}
{Jones}, D.~O., {French}, K.~D., {Agnello}, A., {et~al.} 2023, Transient Name Server Discovery Report, 2023-1228, 1

\bibitem[{{Leonard} \& {Filippenko}(2001)}]{Leonard_2001_typeII}
{Leonard}, D.~C., \& {Filippenko}, A.~V. 2001, \pasp, 113, 920, \dodoi{10.1086/322151}

\bibitem[{{Leonard} {et~al.}(2001){Leonard}, {Filippenko}, {Ardila}, \& {Brotherton}}]{Leonard_2001}
{Leonard}, D.~C., {Filippenko}, A.~V., {Ardila}, D.~R., \& {Brotherton}, M.~S. 2001, \apj, 553, 861, \dodoi{10.1086/320959}

\bibitem[{{Leonard} {et~al.}(2000{\natexlab{a}}){Leonard}, {Filippenko}, {Barth}, \& {Matheson}}]{Leonard_2000}
{Leonard}, D.~C., {Filippenko}, A.~V., {Barth}, A.~J., \& {Matheson}, T. 2000{\natexlab{a}}, \apj, 536, 239, \dodoi{10.1086/308910}

\bibitem[{{Leonard} {et~al.}(2000{\natexlab{b}}){Leonard}, {Filippenko}, {Barth}, \& {Matheson}}]{Leonard_2000_halpha_isp}
---. 2000{\natexlab{b}}, \apj, 536, 239, \dodoi{10.1086/308910}

\bibitem[{{Leonard} {et~al.}(2006){Leonard}, {Filippenko}, {Ganeshalingam}, {Serduke}, {Li}, {Swift}, {Gal-Yam}, {Foley}, {Fox}, {Park}, {Hoffman}, \& {Wong}}]{Leonard_2006}
{Leonard}, D.~C., {Filippenko}, A.~V., {Ganeshalingam}, M., {et~al.} 2006, \nat, 440, 505, \dodoi{10.1038/nature04558}

\bibitem[{{Leonard} {et~al.}(2021){Leonard}, {Dessart}, {Hillier}, {Pignata}, {Williams}, {Hoffman}, {Milne}, {Smith}, {Smith}, \& {Khandrika}}]{Leonard_2021}
{Leonard}, D.~C., {Dessart}, L., {Hillier}, D.~J., {et~al.} 2021, \apjl, 921, L35, \dodoi{10.3847/2041-8213/ac31bf}

\bibitem[{{Li} {et~al.}(2024){Li}, {Hu}, {Li}, {Yang}, {Wang}, {Yan}, {Hu}, {Zhang}, {Mao}, {Riise}, {Gao}, {Sun}, {Liu}, {Xiong}, {Wang}, {Mo}, {Iskandar}, {Xi}, {Xiang}, {Wang}, {Sun}, {Zhang}, {Chen}, {Lin}, {Guo}, {Liu}, {Cai}, {Zhou}, {Zhao}, {Chen}, {Zheng}, {Li}, {Zhang}, {Xu}, {Lyu}, {Castro-Tirado}, {Chufarin}, {Potapov}, {Ionov}, {Korotkiy}, {Nazarov}, {Sokolovsky}, {Hamann}, \& {Herman}}]{Li_2024_23ixf}
{Li}, G., {Hu}, M., {Li}, W., {et~al.} 2024, \nat, 627, 754, \dodoi{10.1038/s41586-023-06843-6}

\bibitem[{{Li} {et~al.}(2011){Li}, {Leaman}, {Chornock}, {Filippenko}, {Poznanski}, {Ganeshalingam}, {Wang}, {Modjaz}, {Jha}, {Foley}, \& {Smith}}]{Li_2011}
{Li}, W., {Leaman}, J., {Chornock}, R., {et~al.} 2011, \mnras, 412, 1441, \dodoi{10.1111/j.1365-2966.2011.18160.x}

\bibitem[{{Mauerhan} {et~al.}(2014){Mauerhan}, {Williams}, {Smith}, {Smith}, {Filippenko}, {Hoffman}, {Milne}, {Leonard}, {Clubb}, {Fox}, \& {Kelly}}]{Mauerhan_2014}
{Mauerhan}, J., {Williams}, G.~G., {Smith}, N., {et~al.} 2014, \mnras, 442, 1166, \dodoi{10.1093/mnras/stu730}

\bibitem[{{Maund}(2024)}]{Maund_2024}
{Maund}, J.~R. 2024, \mnras, 528, 3875, \dodoi{10.1093/mnras/stad2572}

\bibitem[{{Maund} {et~al.}(2023){Maund}, {Wiersema}, {Shrestha}, {Steele}, \& {Hume}}]{Maund_2023}
{Maund}, J.~R., {Wiersema}, K., {Shrestha}, M., {Steele}, I., \& {Hume}, G. 2023, Transient Name Server AstroNote, 135, 1

\bibitem[{{McCall}(1984)}]{McCall_1984}
{McCall}, M.~L. 1984, \mnras, 210, 829, \dodoi{10.1093/mnras/210.4.829}

\bibitem[{{Milne} {et~al.}(2017){Milne}, {Williams}, {Porter}, {Smith}, {Smith}, {Leising}, {Jannuzi}, \& {Green}}]{Milne_2017}
{Milne}, P.~A., {Williams}, G.~G., {Porter}, A., {et~al.} 2017, \apj, 835, 100, \dodoi{10.3847/1538-4357/835/1/100}

\bibitem[{{Moriya} \& {Singh}(2024)}]{Moriya_2024}
{Moriya}, T.~J., \& {Singh}, A. 2024, arXiv e-prints, arXiv:2406.00928, \dodoi{10.48550/arXiv.2406.00928}

\bibitem[{{Moriya} \& {Tominaga}(2012)}]{moriya_2012}
{Moriya}, T.~J., \& {Tominaga}, N. 2012, \apj, 747, 118, \dodoi{10.1088/0004-637X/747/2/118}

\bibitem[{{Morozova} {et~al.}(2017){Morozova}, {Piro}, \& {Valenti}}]{Morozova_2017}
{Morozova}, V., {Piro}, A.~L., \& {Valenti}, S. 2017, \apj, 838, 28, \dodoi{10.3847/1538-4357/aa6251}

\bibitem[{{Morozova} {et~al.}(2018){Morozova}, {Piro}, \& {Valenti}}]{Morozova_2018}
---. 2018, \apj, 858, 15, \dodoi{10.3847/1538-4357/aab9a6}

\bibitem[{{Nagao} {et~al.}(2019){Nagao}, {Cikota}, {Patat}, {Taubenberger}, {Bulla}, {Faran}, {Sand}, {Valenti}, {Andrews}, \& {Reichart}}]{Nagao_2019}
{Nagao}, T., {Cikota}, A., {Patat}, F., {et~al.} 2019, \mnras, 489, L69, \dodoi{10.1093/mnrasl/slz119}

\bibitem[{{Nagao} {et~al.}(2021){Nagao}, {Patat}, {Taubenberger}, {Baade}, {Faran}, {Cikota}, {Sand}, {Bulla}, {Kuncarayakti}, {Maund}, {Tartaglia}, {Valenti}, \& {Reichart}}]{Nagao_2021}
{Nagao}, T., {Patat}, F., {Taubenberger}, S., {et~al.} 2021, \mnras, 505, 3664, \dodoi{10.1093/mnras/stab1582}

\bibitem[{{Nagao} {et~al.}(2023){Nagao}, {Patat}, {Cikota}, {Baade}, {Mattila}, {Kotak}, {Kuncarayakti}, {Bulla}, \& {Ayala}}]{Nagao_2023}
{Nagao}, T., {Patat}, F., {Cikota}, A., {et~al.} 2023, arXiv e-prints, arXiv:2308.00996, \dodoi{10.48550/arXiv.2308.00996}

\bibitem[{{Perley} {et~al.}(2023){Perley}, {Gal-Yam}, {Irani}, \& {Zimmerman}}]{Perlely_2023_23ixfClassification}
{Perley}, D.~A., {Gal-Yam}, A., {Irani}, I., \& {Zimmerman}, E. 2023, Transient Name Server AstroNote, 119, 1

\bibitem[{{Plaszczynski} {et~al.}(2014){Plaszczynski}, {Montier}, {Levrier}, \& {Tristram}}]{Plaszczynski_2014}
{Plaszczynski}, S., {Montier}, L., {Levrier}, F., \& {Tristram}, M. 2014, \mnras, 439, 4048, \dodoi{10.1093/mnras/stu270}

\bibitem[{{Poznanski} {et~al.}(2012){Poznanski}, {Prochaska}, \& {Bloom}}]{Poznanski_2012}
{Poznanski}, D., {Prochaska}, J.~X., \& {Bloom}, J.~S. 2012, \mnras, 426, 1465, \dodoi{10.1111/j.1365-2966.2012.21796.x}

\bibitem[{{Price-Whelan} {et~al.}(2018){Price-Whelan}, {Sip{\H{o}}cz}, {G{\"u}nther}, {Lim}, {Crawford}, {Conseil}, {Shupe}, {Craig}, {Dencheva}, {Ginsburg}, {VanderPlas}, {Bradley}, {P{\'e}rez-Su{\'a}rez}, {de Val-Borro}, {Paper Contributors}, {Aldcroft}, {Cruz}, {Robitaille}, {Tollerud}, {Coordination Committee}, {Ardelean}, {Babej}, {Bach}, {Bachetti}, {Bakanov}, {Bamford}, {Barentsen}, {Barmby}, {Baumbach}, {Berry}, {Biscani}, {Boquien}, {Bostroem}, {Bouma}, {Brammer}, {Bray}, {Breytenbach}, {Buddelmeijer}, {Burke}, {Calderone}, {Cano Rodr{\'\i}guez}, {Cara}, {Cardoso}, {Cheedella}, {Copin}, {Corrales}, {Crichton}, {D{\textquoteright}Avella}, {Deil}, {Depagne}, {Dietrich}, {Donath}, {Droettboom}, {Earl}, {Erben}, {Fabbro}, {Ferreira}, {Finethy}, {Fox}, {Garrison}, {Gibbons}, {Goldstein}, {Gommers}, {Greco}, {Greenfield}, {Groener}, {Grollier}, {Hagen}, {Hirst}, {Homeier}, {Horton}, {Hosseinzadeh}, {Hu}, {Hunkeler}, {Ivezi{\'c}}, {Jain}, {Jenness}, {Kanarek}, {Kendrew}, {Kern}, {Kerzendorf}, {Khvalko},
  {King}, {Kirkby}, {Kulkarni}, {Kumar}, {Lee}, {Lenz}, {Littlefair}, {Ma}, {Macleod}, {Mastropietro}, {McCully}, {Montagnac}, {Morris}, {Mueller}, {Mumford}, {Muna}, {Murphy}, {Nelson}, {Nguyen}, {Ninan}, {N{\"o}the}, {Ogaz}, {Oh}, {Parejko}, {Parley}, {Pascual}, {Patil}, {Patil}, {Plunkett}, {Prochaska}, {Rastogi}, {Reddy Janga}, {Sabater}, {Sakurikar}, {Seifert}, {Sherbert}, {Sherwood-Taylor}, {Shih}, {Sick}, {Silbiger}, {Singanamalla}, {Singer}, {Sladen}, {Sooley}, {Sornarajah}, {Streicher}, {Teuben}, {Thomas}, {Tremblay}, {Turner}, {Terr{\'o}n}, {van Kerkwijk}, {de la Vega}, {Watkins}, {Weaver}, {Whitmore}, {Woillez}, {Zabalza}, \& {Contributors}}]{astropy:2018}
{Price-Whelan}, A.~M., {Sip{\H{o}}cz}, B.~M., {G{\"u}nther}, H.~M., {et~al.} 2018, \aj, 156, 123, \dodoi{10.3847/1538-3881/aabc4f}

\bibitem[{{Riess} {et~al.}(2022){Riess}, {Yuan}, {Macri}, {Scolnic}, {Brout}, {Casertano}, {Jones}, {Murakami}, {Anand}, {Breuval}, {Brink}, {Filippenko}, {Hoffmann}, {Jha}, {D'arcy Kenworthy}, {Mackenty}, {Stahl}, \& {Zheng}}]{Riess_2022}
{Riess}, A.~G., {Yuan}, W., {Macri}, L.~M., {et~al.} 2022, \apjl, 934, L7, \dodoi{10.3847/2041-8213/ac5c5b}

\bibitem[{{Schlafly} \& {Finkbeiner}(2011)}]{Schlafly_2011}
{Schlafly}, E.~F., \& {Finkbeiner}, D.~P. 2011, \apj, 737, 103, \dodoi{10.1088/0004-637X/737/2/103}

\bibitem[{{Schmidt} {et~al.}(1992){Schmidt}, {Stockman}, \& {Smith}}]{Schmidt_1992b}
{Schmidt}, G.~D., {Stockman}, H.~S., \& {Smith}, P.~S. 1992, \apjl, 398, L57, \dodoi{10.1086/186576}

\bibitem[{{Serkowski} {et~al.}(1975){Serkowski}, {Mathewson}, \& {Ford}}]{Serkowski_1975}
{Serkowski}, K., {Mathewson}, D.~S., \& {Ford}, V.~L. 1975, \apj, 196, 261, \dodoi{10.1086/153410}

\bibitem[{{Shrestha} {et~al.}(2020){Shrestha}, {Steele}, {Piascik}, {Jermak}, {Smith}, \& {Copperwheat}}]{Shrestha_2020}
{Shrestha}, M., {Steele}, I.~A., {Piascik}, A.~S., {et~al.} 2020, \mnras, 494, 4676, \dodoi{10.1093/mnras/staa1049}

\bibitem[{{Shrestha} {et~al.}(2024){Shrestha}, {Bostroem}, {Sand}, {Hosseinzadeh}, {Andrews}, {Dong}, {Hoang}, {Janzen}, {Pearson}, {Jencson}, {Lundquist}, {Mehta}, {Ravi}, {Meza Retamal}, {Valenti}, {Brown}, {Jha}, {Macrie}, {Hsu}, {Farah}, {Howell}, {McCully}, {Newsome}, {Padilla Gonzalez}, {Pellegrino}, {Terreran}, {Kwok}, {Smith}, {Schwab}, {Martas}, {Munoz}, {Medina}, {Li}, {Diaz}, {Hiramatsu}, {Tucker}, {Wheeler}, {Wang}, {Zhai}, {Zhang}, {Gangopadhyay}, {Yang}, \& {Gutierez}}]{Shrestha_2024_24ggi}
{Shrestha}, M., {Bostroem}, K.~A., {Sand}, D.~J., {et~al.} 2024, arXiv e-prints, arXiv:2405.18490, \dodoi{10.48550/arXiv.2405.18490}

\bibitem[{{Singh} {et~al.}(2024){Singh}, {Teja}, {Moriya}, {Maeda}, {Kawabata}, {Tanaka}, {Imazawa}, {Nakaoka}, {Gangopadhyay}, {Yamanaka}, {Swain}, {Sahu}, {Anupama}, {Kumar}, {Anche}, {Sano}, {Raj}, {Agnihotri}, {Bhalerao}, {Bisht}, {Bisht}, {Belwal}, {Chakrabarti}, {Fujii}, {Nagayama}, {Matsumoto}, {Hamada}, {Kawabata}, {Kumar}, {Kumar}, {Malkan}, {Smith}, {Sakagami}, {Taguchi}, {Tominaga}, \& {Watanabe}}]{Singh_2024_23ixfPol}
{Singh}, A., {Teja}, R.~S., {Moriya}, T.~J., {et~al.} 2024, arXiv e-prints, arXiv:2405.20989, \dodoi{10.48550/arXiv.2405.20989}

\bibitem[{{Skalidis} {et~al.}(2018){Skalidis}, {Panopoulou}, {Tassis}, {Pavlidou}, {Blinov}, {Komis}, \& {Liodakis}}]{Skalidis_2018}
{Skalidis}, R., {Panopoulou}, G.~V., {Tassis}, K., {et~al.} 2018, \aap, 616, A52, \dodoi{10.1051/0004-6361/201832827}

\bibitem[{{Smartt}(2015)}]{Smartt_2015}
{Smartt}, S.~J. 2015, \pasa, 32, e016, \dodoi{10.1017/pasa.2015.17}

\bibitem[{{Smith}(2014)}]{Smith_2014}
{Smith}, N. 2014, \araa, 52, 487, \dodoi{10.1146/annurev-astro-081913-040025}

\bibitem[{{Smith} {et~al.}(2011){Smith}, {Li}, {Filippenko}, \& {Chornock}}]{Smith_2011}
{Smith}, N., {Li}, W., {Filippenko}, A.~V., \& {Chornock}, R. 2011, \mnras, 412, 1522, \dodoi{10.1111/j.1365-2966.2011.17229.x}

\bibitem[{{Smith} {et~al.}(2023){Smith}, {Pearson}, {Sand}, {Ilyin}, {Bostroem}, {Hosseinzadeh}, \& {Shrestha}}]{Smith_2023}
{Smith}, N., {Pearson}, J., {Sand}, D.~J., {et~al.} 2023, \apj, 956, 46, \dodoi{10.3847/1538-4357/acf366}

\bibitem[{{Smith} {et~al.}(2015){Smith}, {Mauerhan}, {Cenko}, {Kasliwal}, {Silverman}, {Filippenko}, {Gal-Yam}, {Clubb}, {Graham}, {Leonard}, {Horst}, {Williams}, {Andrews}, {Kulkarni}, {Nugent}, {Sullivan}, {Maguire}, {Xu}, \& {Ben-Ami}}]{Smith_2015}
{Smith}, N., {Mauerhan}, J.~C., {Cenko}, S.~B., {et~al.} 2015, \mnras, 449, 1876, \dodoi{10.1093/mnras/stv354}

\bibitem[{{Steele} {et~al.}(2004){Steele}, {Smith}, {Rees}, {Baker}, {Bates}, {Bode}, {Bowman}, {Carter}, {Etherton}, {Ford}, {Fraser}, {Gomboc}, {Lett}, {Mansfield}, {Marchant}, {Medrano-Cerda}, {Mottram}, {Raback}, {Scott}, {Tomlinson}, \& {Zamanov}}]{Steele_2004}
{Steele}, I.~A., {Smith}, R.~J., {Rees}, P.~C., {et~al.} 2004, in Society of Photo-Optical Instrumentation Engineers (SPIE) Conference Series, Vol. 5489, Ground-based Telescopes. Edited by Oschmann, Jacobus M., Jr. Proceedings of the SPIE, Volume 5489, pp. 679-692 (2004)., 679--692, \dodoi{10.1117/12.551456}

\bibitem[{{Tody}(1986)}]{iraf1}
{Tody}, D. 1986, in Society of Photo-Optical Instrumentation Engineers (SPIE) Conference Series, Vol. 627, Instrumentation in astronomy VI, ed. D.~L. {Crawford}, 733, \dodoi{10.1117/12.968154}

\bibitem[{{Tody}(1993)}]{iraf2}
{Tody}, D. 1993, in Astronomical Society of the Pacific Conference Series, Vol.~52, Astronomical Data Analysis Software and Systems II, ed. R.~J. {Hanisch}, R.~J.~V. {Brissenden}, \& J.~{Barnes}, 173

\bibitem[{{Valenti} {et~al.}(2016){Valenti}, {Howell}, {Stritzinger}, {Graham}, {Hosseinzadeh}, {Arcavi}, {Bildsten}, {Jerkstrand}, {McCully}, {Pastorello}, {Piro}, {Sand}, {Smartt}, {Terreran}, {Baltay}, {Benetti}, {Brown}, {Filippenko}, {Fraser}, {Rabinowitz}, {Sullivan}, \& {Yuan}}]{Valenti_2016}
{Valenti}, S., {Howell}, D.~A., {Stritzinger}, M.~D., {et~al.} 2016, \mnras, 459, 3939, \dodoi{10.1093/mnras/stw870}

\bibitem[{Van~Dyk(2017)}]{vandyk_2017}
Van~Dyk, S.~D. 2017, Philosophical Transactions of the Royal Society of London Series A, 375, 20160277, \dodoi{10.1098/rsta.2016.0277}

\bibitem[{{Vasylyev} {et~al.}(2023){Vasylyev}, {Yang}, {Filippenko}, {Patra}, {Brink}, {Wang}, {Chornock}, {Margutti}, {Gates}, {Burgasser}, {Karpoor}, {LeBaron}, {Softich}, {Theissen}, {Wiston}, \& {Zheng}}]{Vasylyev_2023}
{Vasylyev}, S.~S., {Yang}, Y., {Filippenko}, A.~V., {et~al.} 2023, \apjl, 955, L37, \dodoi{10.3847/2041-8213/acf1a3}

\bibitem[{{Vasylyev} {et~al.}(2024){Vasylyev}, {Yang}, {Patra}, {Filippenko}, {Baade}, {Brink}, {Hoeflich}, {Maund}, {Patat}, {Wang}, {Wheeler}, \& {Zheng}}]{Vasylyev_2024}
{Vasylyev}, S.~S., {Yang}, Y., {Patra}, K.~C., {et~al.} 2024, \mnras, 527, 3106, \dodoi{10.1093/mnras/stad3352}

\bibitem[{{Virtanen} {et~al.}(2020){Virtanen}, {Gommers}, {Oliphant}, {Haberland}, {Reddy}, {Cournapeau}, {Burovski}, {Peterson}, {Weckesser}, {Bright}, {van der Walt}, {Brett}, {Wilson}, {Millman}, {Mayorov}, {Nelson}, {Jones}, {Kern}, {Larson}, {Carey}, {Polat}, {Feng}, {Moore}, {VanderPlas}, {Laxalde}, {Perktold}, {Cimrman}, {Henriksen}, {Quintero}, {Harris}, {Archibald}, {Ribeiro}, {Pedregosa}, {van Mulbregt}, \& {SciPy 1. 0 Contributors}}]{scipy}
{Virtanen}, P., {Gommers}, R., {Oliphant}, T.~E., {et~al.} 2020, Nature Methods, 17, 261, \dodoi{10.1038/s41592-019-0686-2}

\bibitem[{{Wang} \& {Wheeler}(2008)}]{Wang_2008}
{Wang}, L., \& {Wheeler}, J.~C. 2008, \araa, 46, 433, \dodoi{10.1146/annurev.astro.46.060407.145139}

\bibitem[{{Wilking} {et~al.}(1982){Wilking}, {Lebofsky}, \& {Rieke}}]{Wilking_1982}
{Wilking}, B.~A., {Lebofsky}, M.~J., \& {Rieke}, G.~H. 1982, \aj, 87, 695, \dodoi{10.1086/113147}

\bibitem[{{Yaron} {et~al.}(2017){Yaron}, {Perley}, {Gal-Yam}, {Groh}, {Horesh}, {Ofek}, {Kulkarni}, {Sollerman}, {Fransson}, {Rubin}, {Szabo}, {Sapir}, {Taddia}, {Cenko}, {Valenti}, {Arcavi}, {Howell}, {Kasliwal}, {Vreeswijk}, {Khazov}, {Fox}, {Cao}, {Gnat}, {Kelly}, {Nugent}, {Filippenko}, {Laher}, {Wozniak}, {Lee}, {Rebbapragada}, {Maguire}, {Sullivan}, \& {Soumagnac}}]{Yaron_2017}
{Yaron}, O., {Perley}, D.~A., {Gal-Yam}, A., {et~al.} 2017, Nature Physics, 13, 510, \dodoi{10.1038/nphys4025}

\bibitem[{{Zhang} {et~al.}(2023){Zhang}, {Lin}, {Wang}, {Zhao}, {Li}, {Liu}, {Yan}, {Xiang}, {Wang}, \& {Bai}}]{Zhang_2023_23ixf}
{Zhang}, J., {Lin}, H., {Wang}, X., {et~al.} 2023, Science Bulletin, 68, 2548, \dodoi{10.1016/j.scib.2023.09.015}

\bibitem[{{Zimmerman} {et~al.}(2024){Zimmerman}, {Irani}, {Chen}, {Gal-Yam}, {Schulze}, {Perley}, {Sollerman}, {Filippenko}, {Shenar}, {Yaron}, {Shahaf}, {Bruch}, {Ofek}, {De Cia}, {Brink}, {Yang}, {Vasylyev}, {Ben Ami}, {Aubert}, {Badash}, {Bloom}, {Brown}, {De}, {Dimitriadis}, {Fransson}, {Fremling}, {Hinds}, {Horesh}, {Johansson}, {Kasliwal}, {Kulkarni}, {Kushnir}, {Martin}, {Matuzewski}, {McGurk}, {Miller}, {Morag}, {Neil}, {Nugent}, {Post}, {Prusinski}, {Qin}, {Raichoor}, {Riddle}, {Rowe}, {Rusholme}, {Sfaradi}, {Sjoberg}, {Soumagnac}, {Stein}, {Strotjohann}, {Terwel}, {Wasserman}, {Wise}, {Wold}, {Yan}, \& {Zhang}}]{Zimmerman_2024}
{Zimmerman}, E.~A., {Irani}, I., {Chen}, P., {et~al.} 2024, \nat, 627, 759, \dodoi{10.1038/s41586-024-07116-6}

\end{thebibliography}
\bibliographystyle{aasjournal}



\end{document}